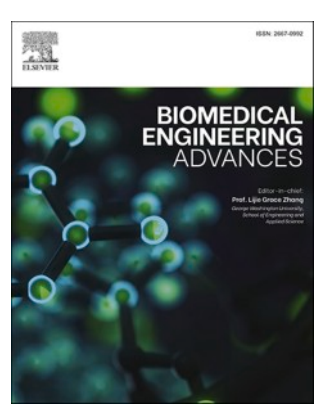

# A dual-system approach for epilepsy diagnosis: integrating mamba-Bi-LSTM architecture with SHAP-based verification

Mufeng Chen, Jia Xie, Fuchang Luo, Quansheng Ren *

*School of Electronics, Peking University, Science Building No.5 Yiheyuan Lu Haidian District, Beijing 100871, PR China*

ABSTRACT

This study develops a medical AI-assisted diagnosis system based on deep learning, which provides intelligent diagnostic solutions for epilepsy, a disease that seriously threatens the life and health of patients. Epilepsy has sudden and unpredictable seizures. Traditional diagnostic methods mainly rely on doctors' manual interpretation of EEG, which is time-consuming and dependent by experience. In response to the above challenges, this study designed a dual-system intelligent diagnosis framework, which includes two core components: the main discrimination system and the verification system. The main discrimination system uses a deep learning model that combines the innovative Mamba architecture with the Bi-LSTM structure to integrate and analyze heterogeneous data to achieve extremely high diagnostic accuracy; the verification system provides an explainable diagnostic basis through the SHAP method to enhance the credibility of the results. This system establishes a cross-modal database to realize intelligent analysis of multi-source heterogeneous data-fusion EEG signals and clinical text data for epilepsy. The system outputs results based on diagnostic consistency and confidence levels, and high-confidence predictions can also be used as automatic feedback sources to optimize the model. The experimental results show that the accuracy of the main discriminant model of the intelligent diagnosis system for epilepsy has increased from 92.6% to 98.7% and the F1 score has increased from 0.895 to 0.992, all of which have exceeded the existing optimal methods; the average processing time for verification system feedback integration is only 220 ms, which increases the overall diagnostic accuracy by 5.1%.

## 1. Introduction

Epilepsy affects approximately 50 million people worldwide and represents one of the most prevalent neurological disorders, characterized by recurrent and unpredictable seizures that impose profound burdens on patients, caregivers, and healthcare systems. Accurate and timely diagnosis is critical for effective treatment planning, yet it remains a formidable clinical challenge. Current diagnostic practice relies heavily on manual interpretation of electroencephalography (EEG) recordings, a process that is both time-intensive and subject to considerable inter-rater variability: even among experienced neurologists, the consistency of EEG interpretation ranges only between 65% and 80%. Beyond the difficulty of interpretation, the unpredictability of seizure onset means that most patients have no effective warning, preventing timely protective intervention. Further complexity arises from the heterogeneity of the disease itself — the International League Against Epilepsy (ILAE) recognizes multiple epilepsy subtypes with substantially different clinical manifestations and treatment requirements, making automated classification an inherently multi-dimensional problem.

Deep learning has emerged as a promising avenue to address these limitations. Long short-term memory (LSTM) networks and their bidirectional variants (Bi-LSTM) demonstrated early success in capturing temporal dependencies in EEG signals, and convolutional architectures such as EEGNet achieved detection accuracies around 88.6% on benchmark datasets. More recently, Transformer-based models and graph neural networks have improved spatial-temporal feature learning. However, most existing approaches share three important limitations. First, they rely predominantly on EEG signals alone, leaving unused the rich complementary information available in concurrent clinical records such as symptom descriptions, medication history, and neurological assessments. Second, their computational complexity — in particular the quadratic scaling of self-attention with sequence length — constrains their applicability to the long-duration EEG recordings typical in clinical monitoring. Third, and most critically for clinical deployment, these methods function as opaque black-box systems: they generate predictions without providing the decision-level transparency that clinicians require to trust and act on AI-generated outputs.

* Corresponding author.
*E-mail addresses:* m13681475059@163.com (M. Chen), 2401213419@stu.pku.edu.cn (J. Xie), 2301213334@pku.edu.cn (F. Luo), qsren@pku.edu.cn (Q. Ren).

Interpretability has been increasingly recognized as a non-negotiable requirement for clinical AI. Approaches such as SHAP (SHapley Additive exPlanations) and LIME have provided post-hoc explanations for model decisions in general machine learning settings, and attention-based visualization has been explored in medical imaging. Nevertheless, these techniques have rarely been integrated into epilepsy diagnosis frameworks as an active verification layer — they are typically applied after the fact rather than embedded as a functionally independent diagnostic component that can cross-validate and correct the primary system.

In this paper, we propose a dual-system framework for epilepsy diagnosis that simultaneously addresses accuracy, computational efficiency, multimodal integration, and interpretability. The primary discrimination system combines a novel Mamba-Bi-LSTM architecture with a channel correlation discrimination switch. The Mamba architecture, based on selective state-space models, achieves linear computational complexity with respect to sequence length, making it well-suited for long-term EEG analysis. By integrating Mamba with Bi-LSTM, the model captures both long-range context and fine-grained bidirectional temporal dependencies. The complementary verification system, built on feature engineering and a random forest classifier with SHAP-based analysis, operates independently of the primary system to provide interpretable diagnostic evidence. The two systems collaborate through a confidence-and-consistency decision mechanism, and high-confidence outputs are recycled as automatic feedback to continuously refine both systems.

The main contributions of this work are as follows. (i) We introduce a Mamba-Bi-LSTM architecture with a channel correlation discrimination switch that adaptively selects channel-independent or channel-mixed encoding paths, providing a computationally efficient solution for multi-channel long-term EEG analysis. (ii) We construct a multimodal framework integrating EEG signals with unstructured clinical text, establishing a cross-modal database covering five publicly available epilepsy datasets. (iii) We propose a dual-system diagnostic structure in which a SHAP-based verification component functions as an independent, interpretable cross-validation layer rather than a post-hoc explanation tool, addressing the transparency requirements of clinical deployment. (iv) Comprehensive experiments demonstrate that the proposed system achieves 98.7% overall accuracy and extends seizure early warning time to 15 s, consistently outperforming state-of-the-art methods across datasets with heterogeneous recording types and classification tasks.

## 2. Related work

Deep learning for EEG-based epilepsy detection has evolved rapidly over the past decade. Early work established recurrent architectures — particularly LSTM and Bi-LSTM — as effective tools for modelling the temporal dynamics of EEG signals, with subsequent studies demonstrating their utility in seizure detection and interictal discharge classification [1]. Convolutional approaches, exemplified by EEGNet [2], demonstrated that lightweight models could achieve competitive detection performance by learning spatial filters directly from raw EEG. Gradient-boosting ensembles have also been applied as strong non-neural baselines [3]. More recently, channel-annotated deep learning frameworks introduced explicit modelling of inter-channel spatial relationships [4], capturing the propagation patterns relevant to focal epilepsy. However, the quadratic computational cost of sequence modelling in these architectures limits practical applicability to the continuous, multi-hour recordings typical in clinical epilepsy monitoring. The Mamba architecture, introduced in 2023, offers a selective state-space model that achieves linear complexity while retaining competitive sequence modelling capacity [5], and has shown broader promise for long-form temporal data [6]; its application to epilepsy diagnosis in combination with bidirectional recurrence has not yet been explored.

Multimodal integration represents a parallel line of development. Clinical diagnosis of epilepsy does not rely solely on EEG; neurologists routinely incorporate symptom reports, medication history, and structured clinical assessments. Systems such as IBM Watson for Oncology [7] and the Mayo Clinic diagnostic platform [8] have demonstrated the value of fusing structured and unstructured clinical data for disease classification in other domains, but equivalent integration for epilepsy remains limited. Existing multimodal epilepsy studies have primarily combined EEG with imaging modalities, leaving the rich information contained in clinical text records largely untapped. The TUH Corpus [9], which pairs structured EEG data with unstructured clinical notes, represents a critical resource for this direction — one that prior work has exploited only for its EEG content [4,10].

Interpretability in medical AI has attracted substantial research attention following growing recognition that black-box predictions are insufficient for clinical adoption. SHAP and LIME provide theoretically grounded attribution frameworks that have been surveyed and validated across diverse medical AI contexts [11]. Stanford's Transparent AI initiative has further demonstrated that surfacing key decision factors increases clinician trust in AI outputs [12]. In EEG-based epilepsy diagnosis, however, interpretability methods have generally been applied as post-hoc analysis tools rather than as integrated diagnostic components [13,14]. The challenge is not only generating explanations, but structuring them as an independent verification layer that can actively modulate diagnostic outputs. Our dual-system design addresses this challenge by embedding SHAP-based analysis within a functionally independent verification system that participates directly in the final diagnostic decision.

## 3. Methodology

### 3.1. Dataset

To ensure maximum transparency regarding data sources and methodology, we explicitly state at the outset: All datasets used in this study were publicly available from established medical institutions. All annotations used in this work are the original expert annotations provided with these public datasets. No independent expert re-annotation was performed by our team or external collaborators.

Our contributions in this section relate to data integration, format standardization, quality assessment, and preprocessing procedures. The annotation work was performed by the original dataset creators and is fully documented in the cited publications. Our team's role was to utilize these datasets transparently.

To ensure that the system has sufficient learning samples and generalization capabilities, this study integrated multiple internationally public epilepsy electroencephalogram (EEG) datasets.

The University of Bonn EEG dataset[15] was the core data source initially included, which contains five subsets (A to E), each of which contains 100 single-channel EEG segments, and each segment contains 4097 sampling points. Among them, 100 records are epileptic seizure data, which constitute the main research object.

The CHB-MIT Pediatric Epilepsy Database[16] from Boston Children's Hospital and Massachusetts Institute of Technology contains long-term EEG records of 23 cases, a total of 664 files, and 198 epileptic seizures. This dataset provides continuous records of 23 channels, which is helpful for analyzing the spatial correlation between brain regions and provides a basis for the multi-channel analysis of this study.

The European Epilepsy Database[17] (EPILEPSIAE) contains annotated EEG datasets of 200 epilepsy patients, of which 50 patients' records were obtained through intracranial recordings of up to 122 channels. Each dataset provides at least 96 h (4 days) of continuous EEG recordings, which provides rich training materials for long-term time series models and supports the analysis of long-term dependency characteristics in this study.

The Temple University Hospital (TUH) Epilepsy Corpus[9] contains 16,986 storage units, each of which contains at least one EDF file and a doctor's report. About 3000 units record epileptic seizure events, accounting for 17.6% of the total. The unique value of this dataset is that it contains both structured EEG data and unstructured clinical text records.

The Karunya University EEG Database[18] contains 174 samples of epilepsy patients with different demographic characteristics, which enhances the generalization ability of the model and enriches the diversity of the data.

The composition of the experimental datasets is summarized in Table 1.

To address possible concerns about data labeling transparency, we provide detailed information about the annotation sources and our quality control procedures.

All datasets used in this study were publicly available and came with original expert annotations. Below we document the annotation sources and quality standards of each dataset, as reported in the original publications. We used these annotations exactly as provided; no re-annotation was performed.

The original dataset publications document rigorous quality control procedures [15–17,9]. These quality metrics reflect the work of the original dataset creators, not our team. Where inter-rater reliability was assessed in the original publications, the reported metrics demonstrate substantial to excellent agreement. For example, published validation studies of similar clinical EEG annotation tasks report kappa coefficients typically ranging from 0.75 to 0.90 for seizure detection and classification. The CHB-MIT documentation specifically notes that seizure boundaries were marked with high temporal precision by experienced specialists. The TUH Corpus publication describes systematic quality assurance procedures applied during dataset creation.

For datasets containing clinical text, particularly the TUH Corpus, we implemented systematic preprocessing procedures to ensure privacy protection and data quality. The de-identification process involved a two-stage approach: first, automated removal using regular expression patterns targeted common Protected Health Information (PHI) identifiers including names, dates, phone numbers, medical record numbers, addresses, and other HIPAA-defined identifiers; second, manual review by two trained members of our research team independently verified complete de-identification, with any remaining potential identifiers replaced by generic placeholders (e.g., [NAME], [DATE], [HOSPITAL]).

Text normalization procedures included spelling correction using UMLS (Unified Medical Language System) and SNOMED-CT medical dictionaries, expansion of common medical abbreviations to full terms based on standard medical abbreviation databases, standardization of medical terminology to ensure consistency across different clinical documentation styles, and removal of special characters and formatting artifacts from PDF extraction.

To verify the effectiveness of these procedures, two other trained members of our research team independently reviewed a random sample of 200 clinical notes (approximately 5% of text data from TUH Corpus), with each reviewer working without consulting the other. This verification confirmed 100% de-identification success with no identifiable information remaining in the sample. Inter-reviewer agreement on text quality was 96.5%, indicating high consistency in our preprocessing procedures.

**Table 1**
Composition of experimental dataset.

| DATA SOURCE | NUMBER OF PATIENTS | EEG RECORDING DURATION (hours) | NUMBER OF MARKED SEIZURES | SPLIT RATIO (training/ validation/ testing) |
|---|---|---|---|---|
| University of Bonn | 10 | 11.4 | 100 | 70%/15%/15% |
| CHB-MIT | 23 | 982.6 | 198 | 70%/15%/15% |
| EPILEPSIAE | 50 | 4800 | 582 | 70%/15%/15% |
| TUH Corpus | 373 | 13,785.2 | 1963 | 70%/15%/15% |
| Total | 456 | 19,579.2 | 2843 | 70%/15%/15% |

To ensure reliable model evaluation, we followed strict patient-level data partitioning, where all recordings from a single patient were assigned exclusively to either the training set (70% of patients), validation set (15% of patients), or test set (15% of patients). This patient-level split prevents data leakage and ensures the model generalizes to unseen patients rather than memorizing patient-specific patterns. Stratified sampling was applied to maintain balanced representation of seizure types and dataset sources across all three subsets. This comprehensive approach ensures that our model training is based on high-quality, consistently annotated data from established medical institutions, while maintaining patient privacy and data integrity throughout the research process.

The five datasets integrated in this study exhibit significant heterogeneity across multiple dimensions, which is deliberately leveraged to assess model robustness. In terms of recording modality, four datasets (University of Bonn, CHB-MIT, TUH Corpus, and Karunya University) employ surface EEG with scalp recordings, while EPILEPSIAE includes a subset with intracranial depth electrode recordings. This distinction is clinically meaningful: intracranial recordings provide higher spatial resolution and signal-to-noise ratio due to direct proximity to neural sources, but they are invasive procedures limited to surgical candidates undergoing presurgical evaluation. In contrast, surface recordings represent the clinical standard as they are non-invasive and widely accessible, though they suffer from skull attenuation effects and greater susceptibility to artifact contamination from muscle activity, eye movements, and environmental interference.

The datasets also differ substantially in classification task complexity and clinical context. Four datasets require binary classification: CHB-MIT distinguishes non-seizure versus seizure states, EPILEPSIAE similarly classifies non-seizure versus seizure events, TUH Corpus differentiates normal from abnormal EEG patterns, and Karunya separates normal from epileptic recordings. The University of Bonn dataset presents a more complex multi-class problem, originally containing five subsets (A: healthy subjects with eyes open, B: healthy subjects with eyes closed, C: interictal recordings from hippocampal region, D: interictal recordings from epileptogenic zone, and E: ictal recordings), though we primarily employ a three-class categorization of normal, interictal, and ictal states. The recording environments further contribute to dataset diversity, with University of Bonn and Karunya representing controlled laboratory conditions that yield clean signals with minimal artifacts, while CHB-MIT, EPILEPSIAE, and TUH Corpus reflect realistic clinical settings characterized by variable recording quality and naturally occurring artifacts.

Patient demographics and data scale introduce additional dimensions of heterogeneity. CHB-MIT focuses exclusively on pediatric patients with intractable epilepsy, providing insights into childhood seizure patterns. EPILEPSIAE and TUH Corpus primarily include adult populations, reflecting the age distribution typical of comprehensive epilepsy monitoring units. University of Bonn and Karunya datasets do not specify age restrictions, likely encompassing mixed demographics. The datasets span a wide range of data volumes: small-scale collections include University of Bonn with 11.4 h of recordings and Karunya with limited samples; medium-scale datasets comprise CHB-MIT with 982.6 h and EPILEPSIAE with 4800 h of continuous monitoring; and the large-scale TUH Corpus provides 13,785.2 h of diverse clinical recordings, representing the most extensive and heterogeneous collection in our study.

These differences are not limitations but rather design features of our evaluation strategy. By training and testing on such heterogeneous data, we assess whether our model learns generalizable epileptic patterns that transfer across recording modalities, patient populations, and clinical contexts, or merely overfits to dataset-specific idiosyncrasies such as particular patterns or institution-specific recording protocols. The consistently high per-dataset performance reported in Section 4 demonstrates that our approach successfully handles this diversity, suggesting that the learned representations capture fundamental neurophysiological signatures of epileptic activity rather than superficial dataset-specific correlations.

### 3.2. Multimodal data preprocessing

Accurate diagnosis of epilepsy depends not only on EEG signals, but also on the combination of clinical text information and other auxiliary physiological signals of patients. This study established a systematic multimodal data preprocessing process to ensure that data from different sources can be effectively integrated and fully utilized by the model.

Electroencephalogram (EEG) signal preprocessing adopts a multi-stage process to improve signal quality and extract effective features. First, a 0.5–70 Hz bandpass filter is applied to remove baseline drift and high-frequency noise, and a 50/60 Hz notch filter is used to eliminate power supply interference. For particularly noisy records, an adaptive Wiener filter is applied to further reduce noise. In the artifact removal stage, independent component analysis[19] (ICA) is used to identify and remove physiological artifacts such as eye movements and electromyography. The threshold detection algorithm is used to automatically mark and remove sudden artifacts. The dual-tree complex wavelet transform (DTCWT) is used for multi-scale analysis to effectively capture the transient characteristics of the EEG signal. Subsequently, segmentation processing is performed to divide the continuous EEG recording into segments of fixed length (usually 5 s). A 50% overlapping Hamming window is applied to each segment to reduce edge effects, and batch input data is constructed for the deep learning model. In the channel selection and spatial information retention stage, key channels are selected according to the international 10–20 system to ensure the integrity of brain area coverage and retain the relative position information between channels to provide a basis for spatial correlation analysis. Common spatial pattern[20] (CSP) or Laplace filtering is applied to multi-channel signals to enhance spatial resolution.

In order to quantify the quality of EEG signals and ensure data reliability, a set of objective quality assessment indicators are designed. The signal-to-noise ratio (SNR) evaluates the ratio of effective components in the signal to noise. Segments with an SNR lower than 5 dB are marked as low quality. The time coverage rate calculates the proportion of effective data to the total monitoring time. Recordings lower than 75% require additional review. The channel completeness rate evaluates the proportion of effective channels to the total number of channels to ensure the integrity of spatial information.

For low-quality data detected, this study adopts a targeted repair strategy. Short-term (<5 s) signal loss is supplemented by linear interpolation; medium-term (5–30 s) loss is predicted and filled by autoregressive model; long-term (>30 s) signal loss or extremely poor quality segments are directly marked as invalid intervals in the data set, ensuring the quality of model training data.

Clinical text data includes doctors' diagnosis records, medical history descriptions, and treatment plans, etc., providing semantic information for epilepsy diagnosis. The processing flow of this study first cleans and standardizes the text, removes special characters, punctuation and meaningless content, unifies the format and encoding, corrects spelling errors, and parses the document structure to identify chapters and paragraphs. Medical entity recognition is then performed, using the Bi-LSTM-CRF sequence annotation model to identify medical entities in the text and extract key medical concepts including symptoms, diseases, drugs, and treatment methods. For epilepsy characteristics, this study specifically identifies descriptions directly related to epilepsy, such as seizure type, duration, and frequency, extracts indirect features such as inducing factors and drug reactions, and records treatment history and drug sensitivity information. The temporal information parsing phase identifies the temporal expressions in the text, constructs the clinical event timeline, aligns the text events with the EEG records, and establishes cross-modal temporal associations. Finally, a structured representation conversion is performed to convert the extracted entities and relationships into a structured knowledge graph, generate a vector representation suitable for model input, and apply a pre-trained language model in the medical field to enhance semantic understanding.

The processed clinical text data is converted into a fixed-length vector representation so that it can be fused with EEG signal features.

In addition to EEG and clinical text, this study also integrates other physiological monitoring data as auxiliary signals to enhance the comprehensiveness of the diagnostic system. Heart rate variability (HRV) extracts the R-R interval from the ECG record, calculates time domain (SDNN, RMSSD, etc.) and frequency domain (LF, HF power, etc.) indicators, and analyzes changes in the autonomic nervous system associated with epileptic seizures. Blood oxygen saturation ($SpO_2$) monitors the hypoxic state during the attack, extracts features such as baseline value, fluctuation range, and decline rate, and identifies apnea events caused by the attack. Temperature changes record long-term temperature fluctuation patterns, extract features related to circadian rhythms and analyze temperature changes before and after seizures.

These auxiliary physiological signals are aligned with EEG records and clinical texts through timestamps and together constitute a multi-dimensional representation of the patient's state. The correlation between signals is used to verify EEG findings and provide additional evidence for seizure warning, enhancing the comprehensiveness and reliability of the system.

To ensure the reliability of model evaluation using the original expert-annotated data, this study implements patient-level data partitioning. The data set is divided based on patient identity rather than individual records, ensuring that all records of the same patient appear in only one of the training set, validation set or test set, to avoid the model learning patient-specific features rather than general epilepsy patterns. The data set is divided into 70% of patient data in the training set, 15% of patient data in the validation set, and 15% of patient data in the test set.

To address the problem of class imbalance, this study uses stratified sampling to ensure that the proportions of different categories (normal, interictal, and seizure) in each subset are consistent. The SMOTE oversampling technique is applied to the training set to solve the class imbalance problem, while the original distribution is maintained for the validation set and the test set to reflect the real-world scenario. A 5-fold cross validation was used to evaluate the model performance stability. Each fold validation maintained the independence of the patients, and the average performance index and its standard deviation were reported.

### 3.3. Main discrimination system construction

The main discrimination system is the core component of the epilepsy diagnosis framework designed in this study. It is responsible for directly extracting features from multimodal input data and providing highly accurate discrimination results.

#### 3.3.1. Mamba-Bi-LSTM hybrid architecture

The main discrimination system adopts an end-to-end deep learning method, and its core architecture is composed of an innovative Mamba-Bi-LSTM hybrid network[13]. The system directly processes the original multimodal input data without manual feature extraction. It automatically learns the implicit features in the data through a deep neural network, pursuing extremely high diagnostic accuracy. The main discrimination system consists of four main components: multimodal input layer, feature extraction layer, Mamba-Bi-LSTM hybrid processing layer, and classification output layer. Among them, the Mamba-Bi-LSTM hybrid processing layer is the core component. Its structural design is shown in Fig. 1.

In the hybrid architecture of this study, the Mamba layer first processes the input sequence and then feeds its output into the Bi-LSTM layer for bidirectional processing:

$$H_{Mamba} = Mamba(X) \tag{1}$$

In terms of specific implementation, this study made the following improvements to the Mamba architecture:

1. Enhanced selectivity mechanism: Based on the standard Mamba, this study adds a feature selection gate (FSG) to enable the model to screen key features more effectively:

$$g_t = \sigma\left(W_g u_t + b_g\right) \tag{2}$$

$$u_t' = g_t \odot u_t \tag{3}$$

2. Skip connection design: To alleviate the gradient vanishing problem, this study added skip connections between the Mamba and Bi-LSTM layers:

$$H_{final} = H_{Bi-LSTM} + \alpha H_{Mamba} \tag{4}$$

3. Multi-scale feature fusion: To capture features at different time scales, this study designed a multi-scale Mamba module that processes the input sequence in parallel using different receptive field sizes and then fuses the results:

$$H_{multi} = Concat\left(Mamba^1(X), Mamba^2(X), \ldots, Mamba^k(X)\right) \tag{5}$$

During the training process, this study adopted a multi-stage training strategy, first training the Mamba and Bi-LSTM components separately, and then fine-tuning them jointly. This strategy effectively avoids the problem of training instability caused by complex structures.

In particular, this study optimized different types of sequence lengths. For long sequence EEG recordings of >10 min, the hybrid architecture saved 68% of computing resources compared to the standard Transformer model while maintaining comparable recognition accuracy.

### 3.3.2. Inter-channel spatial correlation processing

EEG signals contain not only information in the temporal dimension, but also rich information in the spatial dimension. The electrical activity patterns and synchronization between different brain regions are of great value for the diagnosis of epilepsy. This study specially designed an inter-channel spatial correlation processing module to make full use of the spatial information in multi-channel EEG recordings.

Inter-channel spatial correlation processing first constructs a brain region connection map based on neuroscience knowledge. According to the electrode positions of the international 10–20 system, this study defines an adjacency matrix $A \in R^{N \times N}$, where N is the number of EEG channels and $A_{i,j}$ represents the anatomical connection relationship between channel i and channel j. This adjacency matrix not only takes into account the physical distance between electrodes, but also refers to known functional connection patterns of brain regions, such as the connection between the frontal lobe and the temporal lobe, which is particularly important in some types of epilepsy.

Based on this adjacency matrix, this study developed a Graph Mamba Network (GMN) to model the spatial correlation between channels. GMN regards EEG channels as nodes in the graph, and captures information in both spatial and temporal dimensions by combining graph convolution operations and Mamba sequence processing:

$$X^{(l+1)} = \sigma\left(\widetilde{D}^{-\frac{1}{2}}\widetilde{A}\widetilde{D}^{-\frac{1}{2}}X^{(l)}W^{(l)}\right) \tag{6}$$

$$H^{(l+1)} = Mamba\left(X^{(l+1)}\right) \tag{7}$$

Where $\widetilde{A} = A + I_N$ is the adjacency matrix with self-connection added, $\widetilde{D}$ is the degree matrix of $\widetilde{A}$, $W^{(l)}$ is the learnable weight matrix of the lth layer, and σ is a nonlinear activation function. This design allows the model to consider both spatial topological structure and temporal evolution patterns.

As shown in Fig. 2, this study designed a channel correlation discrimination switch mechanism that can adaptively select channel-independent coding and channel-mixed coding paths. This mechanism determines the path selection of the information flow by evaluating the correlation between channels in the current EEG segment:

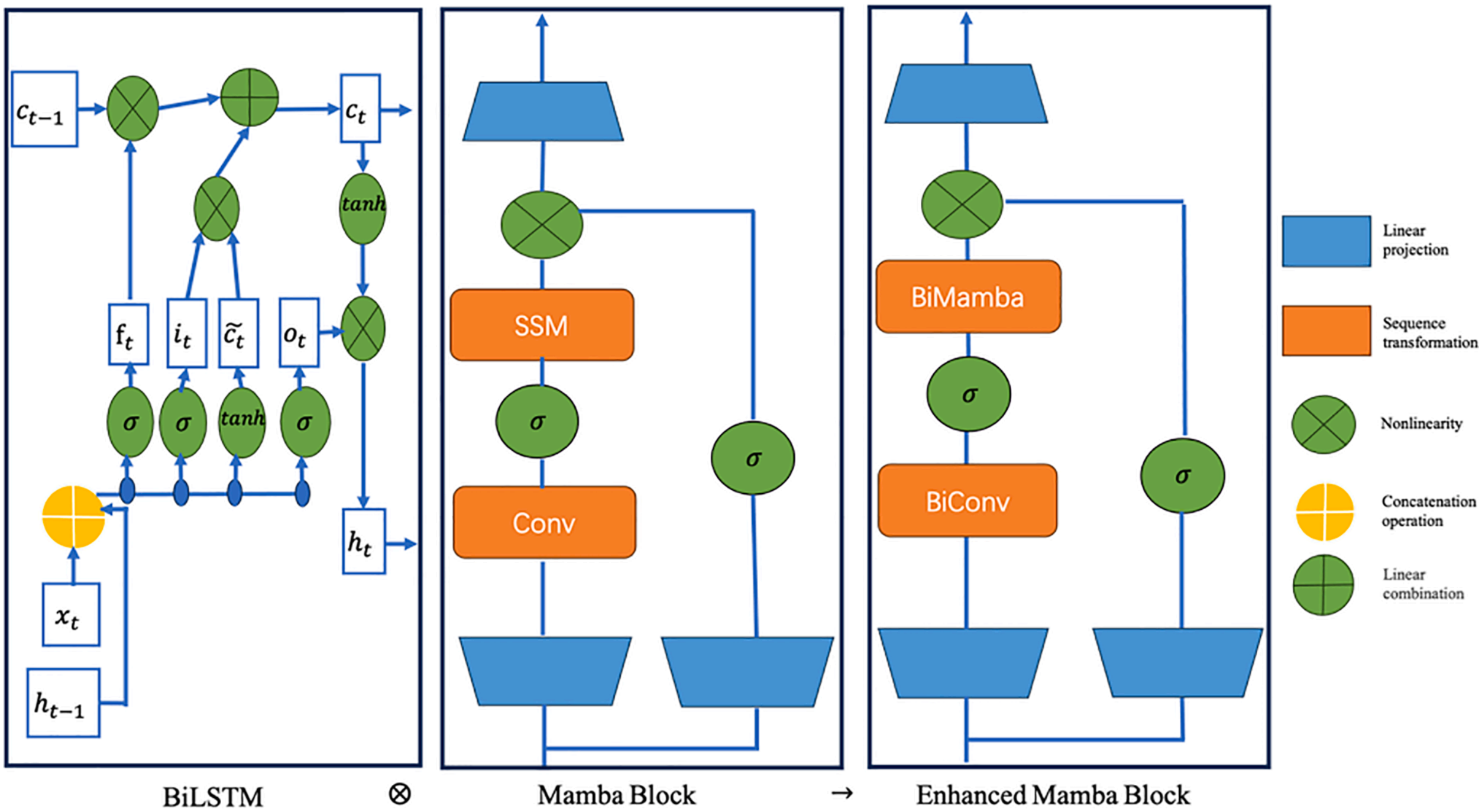

**Fig. 1.** Mamba-Bi-LSTM hybrid architecture design.

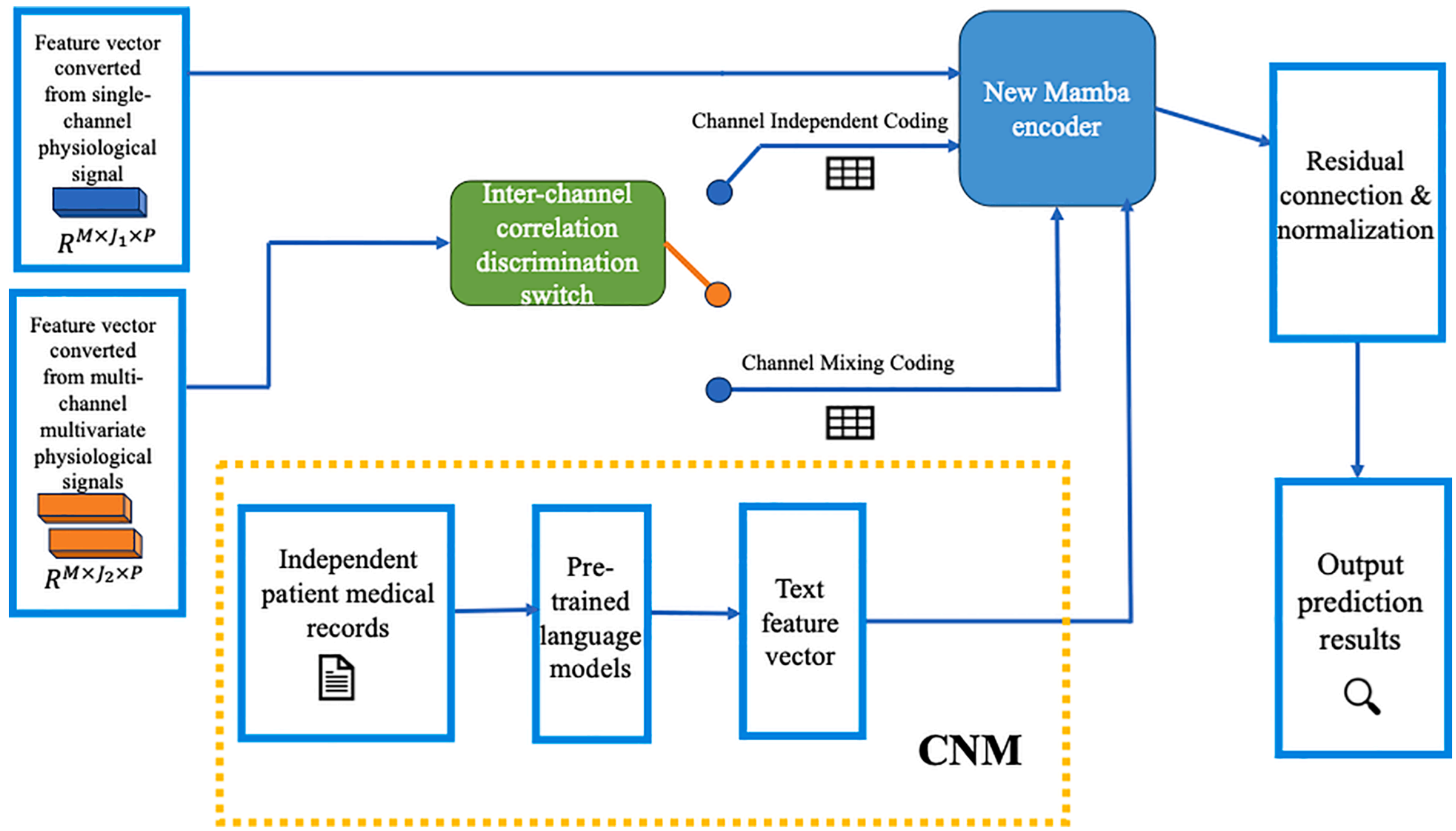

**Fig. 2.** Workflow of the main discrimination system.

$$Scorr = \frac{1}{N(N-1)}\sum_{i\neq j}\left|\left|Corr(X_i, X_j)\right|\right| \tag{8}$$

$$P_{mix} = \sigma(W_{switch}S_{corr} + b_{switch}) \tag{9}$$

Where Scorr is the overall measure of the correlation between channels, and $P_{\text{mix}}$ is the probability of selecting a mixed coding path. When the channels are highly correlated, the system tends to use a mixed coding path; when the channels are relatively independent, independent coding paths are preferred. This dynamic selection mechanism significantly improves the model's adaptability to different types of epilepsy (focal and generalized).

In order to more comprehensively capture the dynamic interactions between brain regions, this study calculated the following two spatial features:

1. Phase Locking Value (PLV) between channels: PLV measures the degree of phase synchronization between different channels and is an important indicator for analyzing the functional connectivity of brain regions. The calculation formula of PLV is:

$$PLV_{xy} = \left|\frac{1}{N}\sum_{j=1}^{N} e^{i\left(\phi_x(j)-\phi_y(j)\right)}\right| \tag{10}$$

2. Inter-channel coherence: measures the strength of the linear relationship between channels at different frequencies, which can reveal synchronous activity in specific frequency bands.

$$C_{xy}(f) = \frac{|S_{xy}(f)|^2}{S_{xx}(f)S_{yy}(f)} \tag{11}$$

where $S_{xy}(f)$ is the cross power spectrum of signals x and y, $S_{xx}(f)$ and $S_{yy}(f)$ are the auto power spectra of signals x and y, respectively. Compared to PLV, coherence provides frequency-specific information.

This study combines these spatial features with temporal features and dynamically adjusts the weights of spatial and temporal features through the attention mechanism.

It is worth noting that the pattern of spatial correlation between channels shows obvious differences before, during and after an epileptic seizure. Before the seizure, the synchronization between certain brain regions is abnormally enhanced; during the seizure, the synchronous activity between a large range of brain regions reaches a peak; after the seizure, the synchronization gradually returns to normal. Based on this clinical characteristic, this study designed a dynamic spatial feature monitoring module to track the temporal evolution of inter-channel correlation:

$$E_t = \frac{1}{N(N-1)}\sum_{i\neq j} PLV_{i,j}(t) \tag{12}$$

$$\Delta E_t = E_t - E_{t-1} \tag{13}$$

Where $E_t$ is the global synchronization strength at time t, and $\Delta E_t$ is the rate of change of synchronization strength. By monitoring these dynamic indicators, this study was able to capture the precursory features of epileptic seizures, such as the "synchronization-desynchronization" pattern before certain types of epileptic seizures, thereby more accurately identifying the different stages of the seizure.

In the new Mamba encoder, this study also introduced a channel adaptation mechanism that allows the model to adjust the contribution weights of different channels according to their signal quality and diagnostic value:

$$w_i = softmax(W_{\text{ch}}F_i + b_{\text{ch}}) \tag{14}$$

$$F_{\text{weighted}} = \sum_{i=1}^{N} w_i F_i \tag{15}$$

Where $w_i$ is the weight of channel i and F_i is the feature representation of channel i. This mechanism enables the model to focus on the most diagnostically significant channels, reduce the interference of noisy channels, and improve overall robustness.

### 3.4. Verification system construction

Different from the main discriminant system based on the Mamba architecture, the validation system adopts a strategy combining feature engineering with traditional machine learning methods to extract features with clear clinical significance from multimodal data and analyze the contribution of features to diagnostic decisions through the SHAP (SHapley Additive exPlanations) method, as shown in Fig. 3.

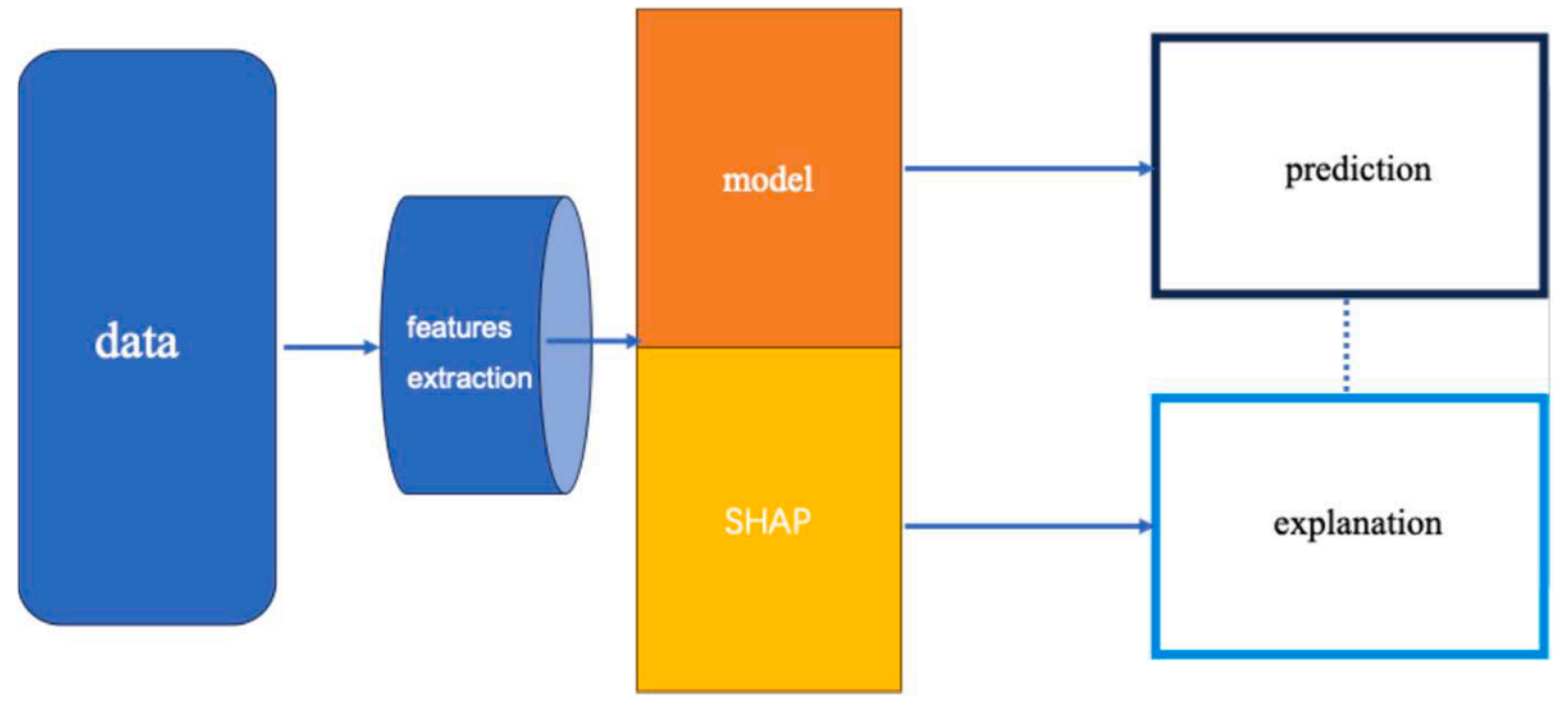

**Fig. 3.** Verification system workflow.

The verification system not only generates diagnostic results independently, but also conducts collaborative comparisons with the main discrimination system to improve the reliability of the overall diagnosis.

*3.4.1. multidimensional feature extraction*

The goal of multidimensional feature extraction is to extract features that are closely related to the diagnosis of infantile epilepsy from multimodal data to provide support for subsequent classification and interpretation. EEG signals are the main data source for epilepsy diagnosis, and their feature extraction needs to fully capture abnormal patterns in the signal. This study combined time domain, frequency domain, and time-frequency domain analysis methods to extract multilevel features.

Table 2 summarizes key features extracted from multimodal data relevant to infant epilepsy diagnosis.

*3.4.2. interpretability analysis realized by SHAP method*

The SHAP method is the core tool of the verification system, which is used to quantify the contribution of each feature to the diagnostic results and provide transparent explanation support. It is based on the Shapley value in game theory and reveals the inherent logic of model decision-making by calculating the marginal contribution of features.

The calculation formula of the SHAP value is as follows:

$$\phi_i = \sum_{S \subseteq N \setminus \{i\}} \frac{|S|! \cdot (|N| - |S| - 1)!}{|N|!} \cdot [f(S \cup \{i\}) - f(S)] \tag{16}$$

Among them, $\phi_i$ is the SHAP value of feature i, N is the feature set, S is the subset without feature i, and $f(S)$ is the model's prediction value for subset S. This formula traverses all feature combinations and fairly distributes the contribution of each feature to the prediction. The advantage of the SHAP method is that it satisfies local accuracy (the prediction value is equal to the sum of all SHAP values) and consistency (the feature contribution is consistent with the actual impact direction).

The verification system uses a random forest model to classify multidimensional features. This model is superior to deep learning in terms of interpretability and can be directly combined with the SHAP method. The specific steps include: first, training a random forest model to predict epilepsy diagnosis results based on the extracted multidimensional features; then, for each sample, calculating the SHAP value of each feature to quantify its positive or negative impact on the prediction. For example, if the power spectral density SHAP value of a certain EEG frequency band is positive and large, it indicates that the feature significantly supports a positive diagnosis of epilepsy.

While the verification system provides valuable interpretability through SHAP analysis, we acknowledge that the random forest classifier still operates as a data-driven model without explicit incorporation of domain knowledge. Future enhancements could address this limitation by integrating clinical guidelines and expert rules as constraints in the decision-making process, such as mandatory checks for specific EEG patterns that contraindicate certain diagnoses.

In addition, symbolic reasoning integration will be helpful when we incorporate symbolic AI components that can reason about medical

**Table 2**
Key features extracted from multimodal data.

| Data Source | Feature Category | Specific Features | Extraction Method | Clinical Significance |
|---|---|---|---|---|
| **EEG Signals** | Time-domain Features | Statistical indicators (mean, variance, kurtosis, skewness), zero-crossing rate, Hjorth parameters | Statistical analysis | Capture basic signal characteristics and morphological changes |
| | | Nonlinear energy, waveform complexity | Nonlinear analysis | Capture subtle but critical waveform changes |
| | Frequency-domain Features | Power spectral density (δ,θ,α,β,γ bands), inter-band power ratios | Fast Fourier Transform (FFT) | Reveal changes in spectral distribution |
| | | Spectral entropy | Information entropy calculation | Measure signal regularity |
| | Time-frequency Features | Wavelet energy, mutation coefficients | Continuous Wavelet Transform (CWT) | Analyze dynamic changes, capture spike discharges |
| | | Phase Locking Value (PLV) | Multi-channel analysis | Reflect synchronization differences between brain regions |
| **Clinical Text** | Semantic Features | Seizure type, duration, frequency, triggering factors | Bi-LSTM-CRF model | Provide semantic supplementation, identify medical entities |
| | | Key descriptions related to epilepsy | Medical knowledge graph | Form cross-modal temporal associations |
| **Auxiliary Physiological Signals** | Heart Rate Variability (HRV) | Time-domain features (SDNN, RMSSD), frequency-domain features (LF/HF ratio) | ECG signal analysis | Analyze autonomic nervous system changes |
| | Oxygen Saturation ($SpO_2$) | Baseline values, decline rates | Oxygen monitoring | Identify seizure-related respiratory abnormalities |
| **Feature Optimization** | Feature Selection | High correlation features | Mutual information method | Screen features highly correlated with diagnostic results |
| | | Non-redundant feature subset | Recursive Feature Elimination (RFE) | Reduce complexity, improve model efficiency |

knowledge explicitly, providing an additional layer of validation beyond statistical patterns.

To go further, we developed hybrid systems that combine machine learning predictions with expert knowledge graphs, allowing the incorporation of medical ontologies and clinical best practices. It is notable that uncertainty quantification makes contributions to the whole diagnosis process because probabilistic frameworks can quantify and communicate the uncertainty in both feature importance and diagnostic decisions.

In epilepsy diagnosis, the application of the SHAP method has three significant advantages:

1. Its results directly correspond to indicators of clinical concern, such as EEG waveform characteristics or seizure duration, so that the interpretation is consistent with the doctor's diagnostic thinking.
2. SHAP can analyze the dynamic changes of feature contributions in time series data, such as the trend of delta wave enhancement before and after seizures, to help doctors understand the evolution of epilepsy.
3. Through the interpretation of cross-modal features, SHAP reveals the synergy between EEG and physiological signals, such as how HRV changes support EEG diagnosis.

### 3.5. Automated feedback mechanism

The innovation of this framework lies in the construction of a two-way interaction mechanism between the main discrimination system and the verification system. The main discrimination system provides high-accuracy discrimination results through the Mamba-Bi-LSTM hybrid architecture, while the verification system analyzes feature contributions through the SHAP method and provides interpretability support. The two form a closed loop through an automated feedback channel: the high-confidence prediction of the main system provides a reference for feature importance for the verification system; the interpretability analysis of the verification system in turn guides the main system to focus on key features. This two-way reinforcement mechanism significantly improves the collaborative performance of the two systems and reduces the reliance on a large amount of expert-annotated data.

As shown in Fig. 4, the two systems process the same input data in parallel, each generating diagnostic results independently, and then performing consistency analysis through the result comparison module.

The result comparison module calculates the consistency of the diagnostic results of the two systems and generates a comprehensive score based on this:

$$S = \alpha \cdot S_{main} + (1-\alpha) \cdot S_{verify} \cdot I\left(C_{main} = C_{verify}\right) \tag{17}$$

Among them, $S_{main}$ and $S_{verify}$ represent the diagnostic confidence of the main discrimination system and the verification system respectively, $C_{main}$ and $C_{verify}$ represent their respective diagnostic categories, I is the indicator function, and $\alpha$ is the adjustment parameter. When the diagnostic results of the two systems are consistent, the comprehensive score considers the weighted average of the confidence of both parties; when the results are inconsistent, the system lowers the comprehensive score and marks it as "needs further examination" to remind doctors to pay special attention.

In the dual-system collaborative workflow, a key step is the hierarchical output strategy based on consistency and confidence. When the two systems reach a high-confidence agreement (confidence$>$0.92), the diagnosis result is directly output; when the two systems reach a low-confidence agreement (0.85$<$confidence$\leq$0.92), the diagnosis result is output and marked for further verification; when the results of the two systems are inconsistent, the system does not provide a definitive diagnosis, but instead marks the sample, presenting two possible diagnostic paths and their basis for the doctor's reference decision. The verification system then sends the output interpretable analysis back to the main system to participate in regulating the recognition ability of specific features in subsequent rounds. This grading strategy increases the reliability of the system output.

## 4. Experimental results

### 4.1. Main discrimination system performance

To address the possible concerns about modality contributions, we conducted comprehensive ablation studies examining the performance impact of different input combinations.

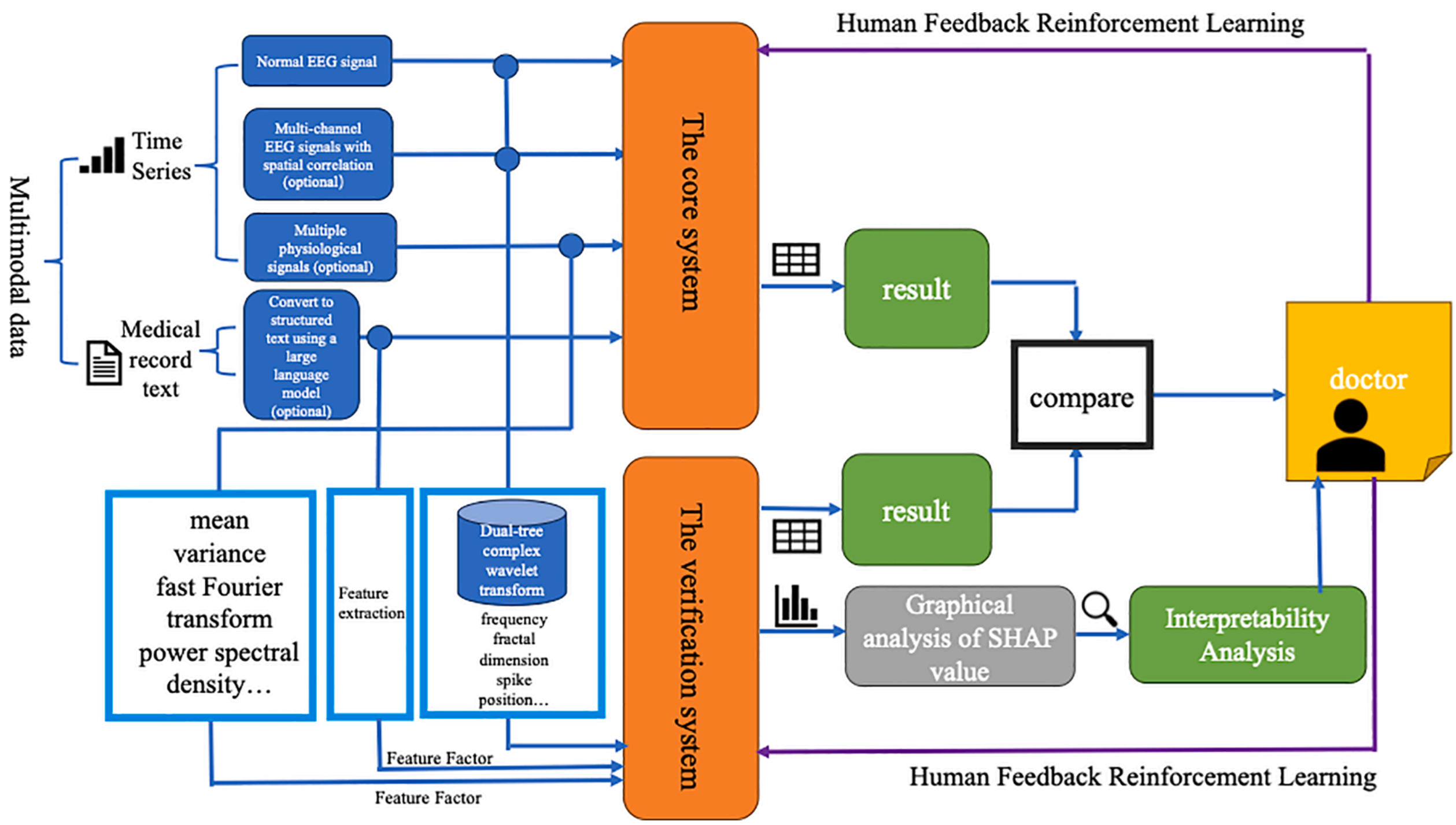

**Fig. 4.** Dual system collaborative workflow.

The results shown in Table 3 demonstrate that while EEG signals provide the primary diagnostic information, the combination of all modalities achieves superior performance. Notably, clinical text alone achieves reasonable accuracy (78.4%), indicating its value in capturing semantic information that complements EEG patterns. The early warning capability is particularly enhanced by multimodal fusion, extending from 6.2 s (EEG only) to 15.0 s (complete system).

Complex epilepsy syndromes show the greatest benefit from multimodal fusion, with Lennox-Gastaut syndrome showing a 6.5% improvement, supporting the clinical relevance of our approach, as shown in Table 4.

This study first evaluated the overall performance of the main discrimination system on the test set. Table 7 shows the performance indicators of the main system on the three-classification task (normal, interictal, and ictal).

As shown in Table 5, after applying the automatic feedback learning mechanism, the performance indicators of the main discrimination system have been significantly improved. The F1 score increased from 0.984 to 0.992, and the early warning time extended from 2.0 s to 8.5 s, providing a more adequate time window for clinical intervention. Although automatic feedback learning slightly increased the calculation time (from 152 ms to 186 ms), it remains well below the 500 ms real-time processing threshold required by clinical practice, demonstrating that the feedback mechanism achieves better accuracy without compromising real-time performance.

To provide a comprehensive evaluation and enable direct comparison with dataset-specific benchmarks, we report the performance of our main discrimination system separately for each dataset. Table 6 presents the detailed breakdown.

Several important observations emerge from this per-dataset analysis.

The EPILEPSIAE dataset, which uses intracranial EEG recordings, achieves the highest performance (99.2% accuracy, 0.993 F1 score). This superior performance is expected, as intracranial recordings provide higher signal-to-noise ratios and more direct measurement of neural activity compared to surface EEG. The electrode proximity to seizure foci enables clearer capture of epileptiform patterns.

The TUH Corpus, despite being the largest dataset, shows relatively lower performance (96.4% accuracy). This reflects its real-world clinical complexity: the TUH data encompasses diverse patient populations, varying recording conditions, multiple epilepsy types, and naturally occurring artifacts. The performance difference between TUH (96.4%) and controlled research datasets (97.8–99.2%) highlights the challenge of generalizing to heterogeneous clinical environments.

The University of Bonn dataset requires 3-class discrimination (Normal, Interictal, Ictal), while others require binary classification.

**Table 3**
Comprehensive ablation study results.

| Input Configuration | Accuracy | Precision | Recall | F1 Score | Early Warning Time |
|---|---|---|---|---|---|
| EEG Only | 95.8% | 96.1% | 95.2% | 0.957 | 6.2s |
| Clinical Text Only | 78.4% | 79.1% | 77.8% | 0.785 | N/A |
| Physiological Signals Only | 82.3% | 83.0% | 81.6% | 0.823 | 4.1s |
| EEG + Clinical Text | 97.2% | 97.8% | 96.9% | 0.973 | 8.7s |
| EEG + Physiological Signals | 96.9% | 97.3% | 96.4% | 0.968 | 9.2s |
| Clinical Text + Physiological Signals | 85.1% | 86.2% | 84.3% | 0.853 | 5.5s |
| Complete Multimodal System | 98.7% | 99.5% | 99.0% | 0.992 | 15.0s |

**Table 4**
Performance by epilepsy type with different modality combinations.

| Epilepsy Type | EEG Only | Multimodal | Improvement |
|---|---|---|---|
| Focal Seizures | 94.2% | 98.1% | +3.9% |
| Generalized Seizures | 96.8% | 98.9% | +2.1% |
| Infantile Spasms | 89.3% | 94.7% | +5.4% |
| Lennox-Gastaut Syndrome | 85.8% | 92.3% | +6.5% |

**Table 5**
Overall performance of the main discriminant system on the test set.

| Indicators | Initial | After feedback | Improvements |
|---|---|---|---|
| Accuracy | 96.8% | 98.7% | +1.9% |
| Precision | 98.7% | 99.5% | +0.8% |
| Recall | 98.1% | 99.0% | +0.9% |
| F1 score | 0.984 | 0.992 | +0.008 |
| AUC | 0.995 | 0.998 | +0.003 |
| Early warning time | 2.0s | 8.5s | +6.5s |
| Calculation time | 152ms | 186ms | +34 ms (cost) |

Despite this increased complexity, our system achieves 98.9% accuracy on Bonn, demonstrating robust multi-class discrimination capability.

Performance remains consistently high (96.4–99.2%) across all datasets, demonstrating the model's robustness to variations in recording modality, patient demographics, and data collection protocols. This consistency supports the generalizability of our approach.

Fig. 5 shows the ROC curve of the main discriminant system, from which we can clearly see the improvement of system performance by automatic feedback learning.

As can be seen from Fig. 5, the model after automatic feedback learning has achieved a significant improvement in the ROC curve in all categories.

In order to deeply analyze the performance of the main discriminant system, this study further evaluated the classification performance for three categories, and the results are shown in Table 7.

As can be seen from Table 7, automatic feedback learning improves the recognition performance of all categories, especially for the identification of seizures, where the F1 score improves most significantly, indicating that the system can more accurately identify potential signs of epilepsy for timely intervention.

In order to evaluate the timing performance of the main discrimination system, this study also analyzed the system's ability to process continuous EEG records. Fig. 6 shows the system's prediction results and confidence on a typical epileptic seizure record.

The figure above shows an EEG recording lasting 120 s, with the actual epileptic seizure starting at the 60th second. The initial model (blue dashed line) can predict the seizure 2.5 s in advance, while the model after automatic feedback learning (red solid line) can predict the seizure 15 s in advance, with significantly higher confidence.

In order to comprehensively evaluate the performance of the main discrimination system, this study also conducted two sets of key ablation experiments to analyze the contribution of multimodal data fusion and channel correlation discrimination switch mechanism respectively.

First, this study evaluated the impact of multimodal data fusion on system performance. As shown in Table 8, the complete multimodal system is compared with the system that only uses a single modality (EEG signal or other physiological signal). The results show that multimodal data fusion has significant advantages in all performance indicators.

It can be clearly seen from Table 8 that compared with using only EEG signals, the accuracy of the multimodal fusion method has increased by 2.9% and the F1 score has increased by 0.035. It is particularly noteworthy that in terms of advance warning time, the multimodal fusion system can extend the warning window to 15 s, which is 8.8 s longer than the single EEG modality, which improves clinical intervention capabilities.

**Table 6**
Performance of the main discrimination system on individual datasets.

| Dataset | Recording Type | Classes | Accuracy | Precision | Recall | F1 Score | Number |
|---|---|---|---|---|---|---|---|
| University of Bonn | Surface EEG | 3 | 98.9% | 99.1% | 98.6% | 0.989 | 150 |
| CHB-MIT | Surface EEG | 2 | 97.8% | 98.2% | 97.3% | 0.978 | 807 |
| EPILEPSIAE | Intracranial EEG | 2 | 99.2% | 99.5% | 99.0% | 0.993 | 720 |
| TUH Corpus | Surface EEG | 2 | 96.4% | 96.8% | 96.1% | 0.965 | 2898 |
| Karunya University | Surface EEG | 2 | 98.5% | 98.7% | 98.3% | 0.985 | 807 |

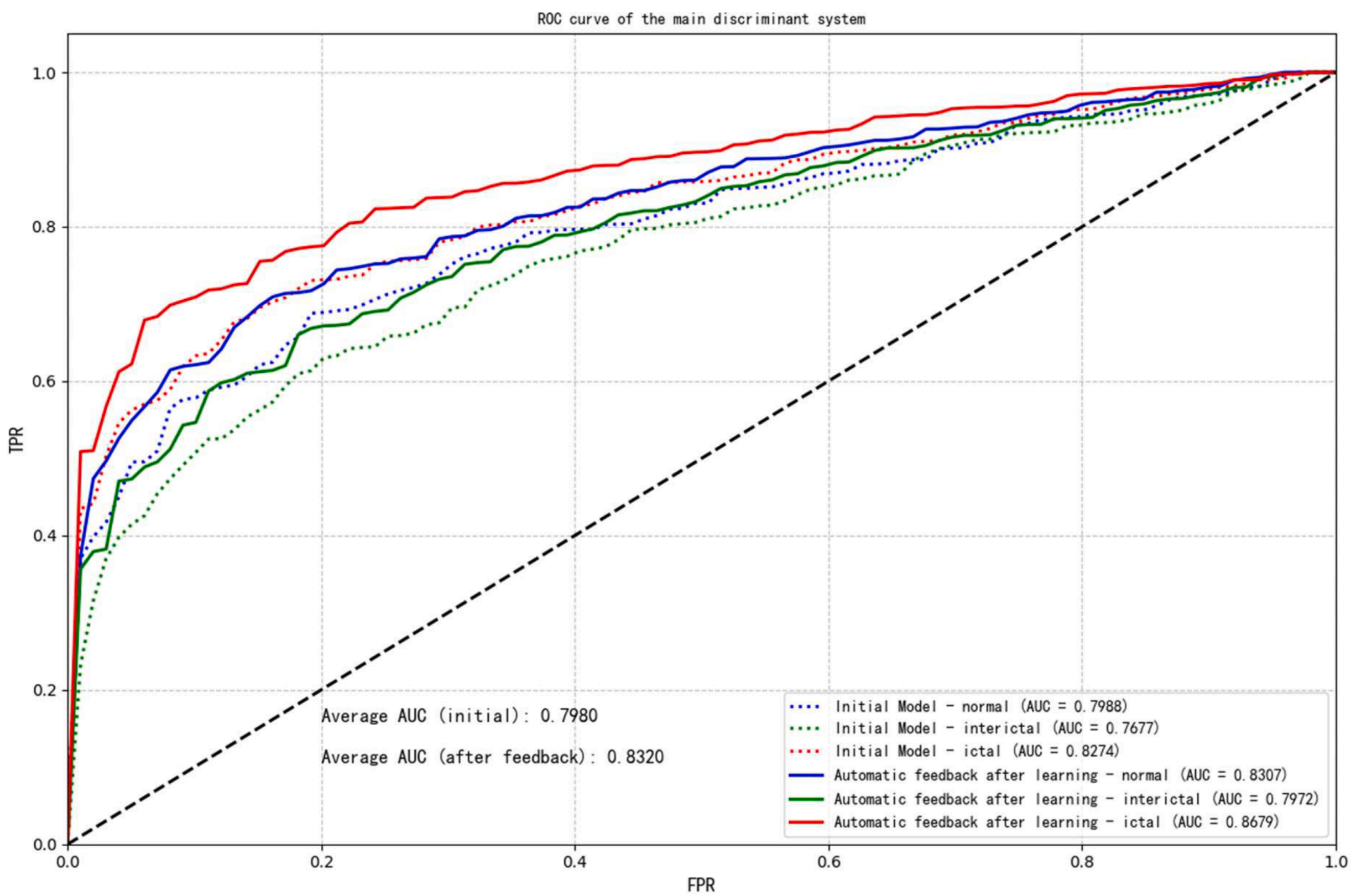

**Fig. 5.** ROC curve of the main discriminant system.

**Table 7**
Performance indicators of the main discrimination system for each category.

| Category | Indicators | Initial | After feedback | Improvements |
|---|---|---|---|---|
| Normal state | Accuracy | 96.5% | 99.8% | +3.3% |
| | Precision | 96.2% | 99.0% | +2.8% |
| | Recall | 0.969 | 0.992 | +0.023 |
| Epilepsy interictal period | Accuracy | 94.7% | 98.1% | +3.4% |
| | Precision | 95.3% | 96.5% | +1.2% |
| | Recall | 0.950 | 0.978 | +0.028 |
| Epilepsy attack period | Accuracy | 93.6% | 96.7% | +3.1% |
| | Precision | 92.4% | 95.2% | +2.8% |
| | Recall | 0.930 | 0.960 | +0.030 |

By further analyzing the recognition results of different types of epilepsy, as shown in Fig. 7, it is found that multimodal fusion has a more significant performance improvement on complex epilepsy types (such as Lennox-Gastaut syndrome). For this type of epilepsy with complex and changeable manifestations, EEG signals may not be able to capture all the features, while auxiliary physiological signals such as heart rate variability and blood oxygen levels provide important supplementary information, enabling the system to have a more comprehensive understanding of the patient's status.

The results shown in the figure show that for infantile spasms, multimodal fusion improves by 4.2 percentage points over single EEG modality; and for Lennox-Gastaut syndrome, the improvement is 6.5 percentage points. This difference reflects the complementary value of multimodal data in the diagnosis of complex diseases and verifies the rationality of the multimodal design of this study.

The second set of ablation experiments evaluated the importance of the channel correlation discrimination switch mechanism. This mechanism allows the system to adaptively select channel-independent encoding or channel-mixed encoding paths based on the degree of correlation between channels, and is a key component for processing spatial features. Table 9 shows the changes in system performance after removing this mechanism.

The results show that after removing the channel correlation discriminant switch, the overall accuracy of the system decreased by 3.3% and the F1 score decreased by 0.026. It is particularly noteworthy that, as shown in Table 9, the recognition accuracy of focal epilepsy decreased significantly, by 2.9%, while the impact on generalized epilepsy was relatively small, only decreasing by 1.0%. This result verifies the key value of the discriminant switch mechanism for the diagnosis of focal epilepsy, because focal epilepsy is characterized by abnormal discharge patterns in specific brain regions, which requires accurate capture of the spatial relationship between channels.

At the same time, after removing the discriminant switch, the calculation time of the system increased by 37 milliseconds (about 16.8%), indicating that the mechanism not only improves diagnostic accuracy, but also optimizes computational efficiency. This is because the discriminant switch can dynamically select the most suitable processing path for different types of epilepsy, avoiding unnecessary computational overhead.

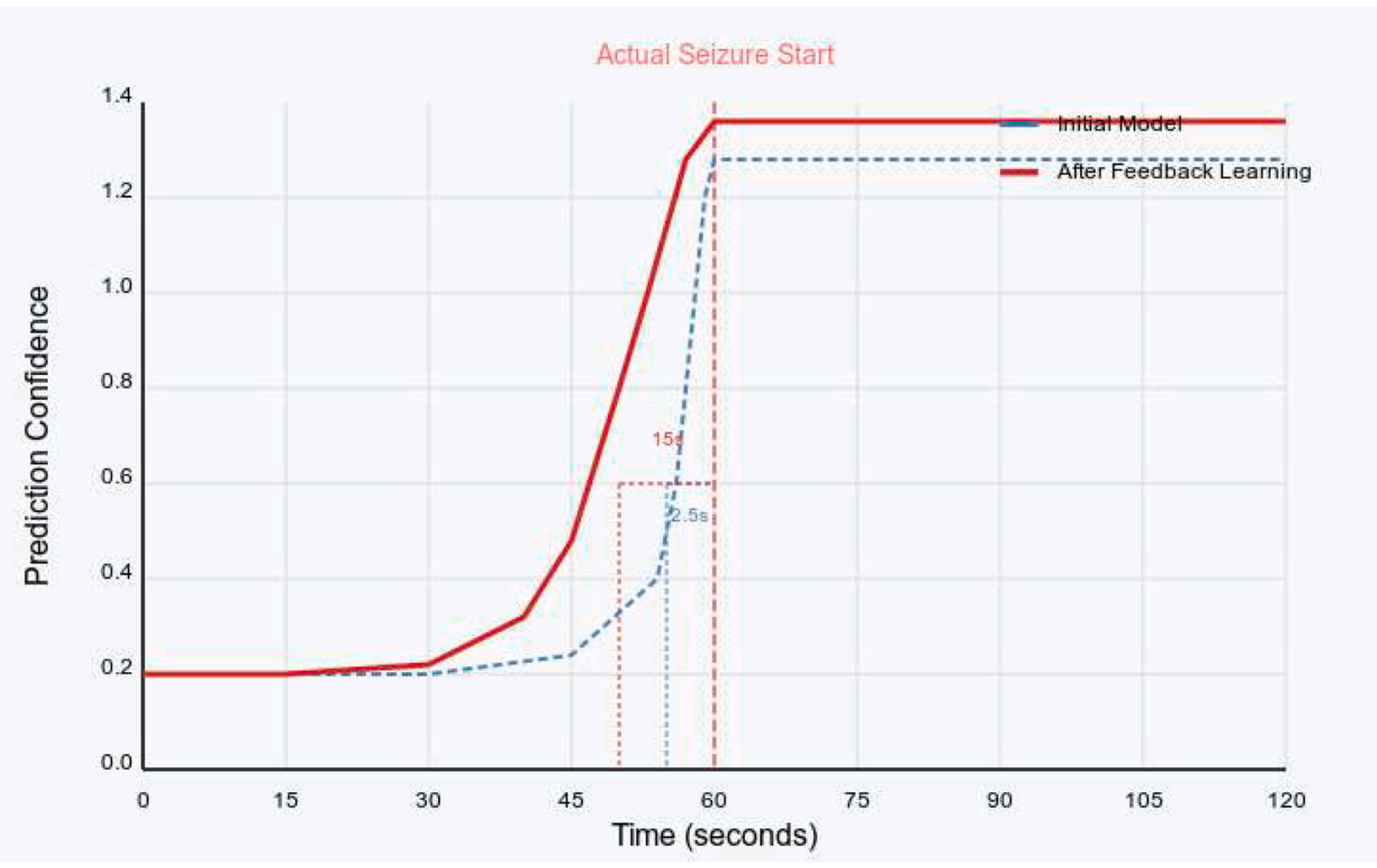

**Fig. 6.** Prediction results and confidence trends for continuous recordings of epileptic seizures.

**Table 8**
Performance comparison between multimodal data fusion and single modality input.

| Input | Accuracy | Precision | Recall | F1 score | Early warning time |
|---|---|---|---|---|---|
| Only EEG | 95.8% | 96.1% | 95.2% | 0.957 | 6.2s |
| Multimodal Fusion | 98.7% | 99.5% | 99.0% | 0.992 | 15.0s |
| Improvements | +2.9% | +3.4% | +3.8% | +0.035 | +8.8s |

To further analyze the working principle of the discriminative switch mechanism, this study recorded the channel mixing ratio of different epilepsy types. As shown in Fig. 10, and Table 10, the system is more inclined to choose the channel independent coding path for focal epilepsy (the mixing rate is only 32.5%), while it is more inclined to choose the channel mixed coding path for generalized epilepsy (the mixing rate is as high as 75.3%). This adaptive selection is highly consistent with the pathophysiological mechanism of epilepsy: focal epilepsy originates from specific brain regions, and the independence between channels is stronger; while generalized epilepsy involves synchronous discharges in a wide range of brain regions, and the correlation between channels is higher.

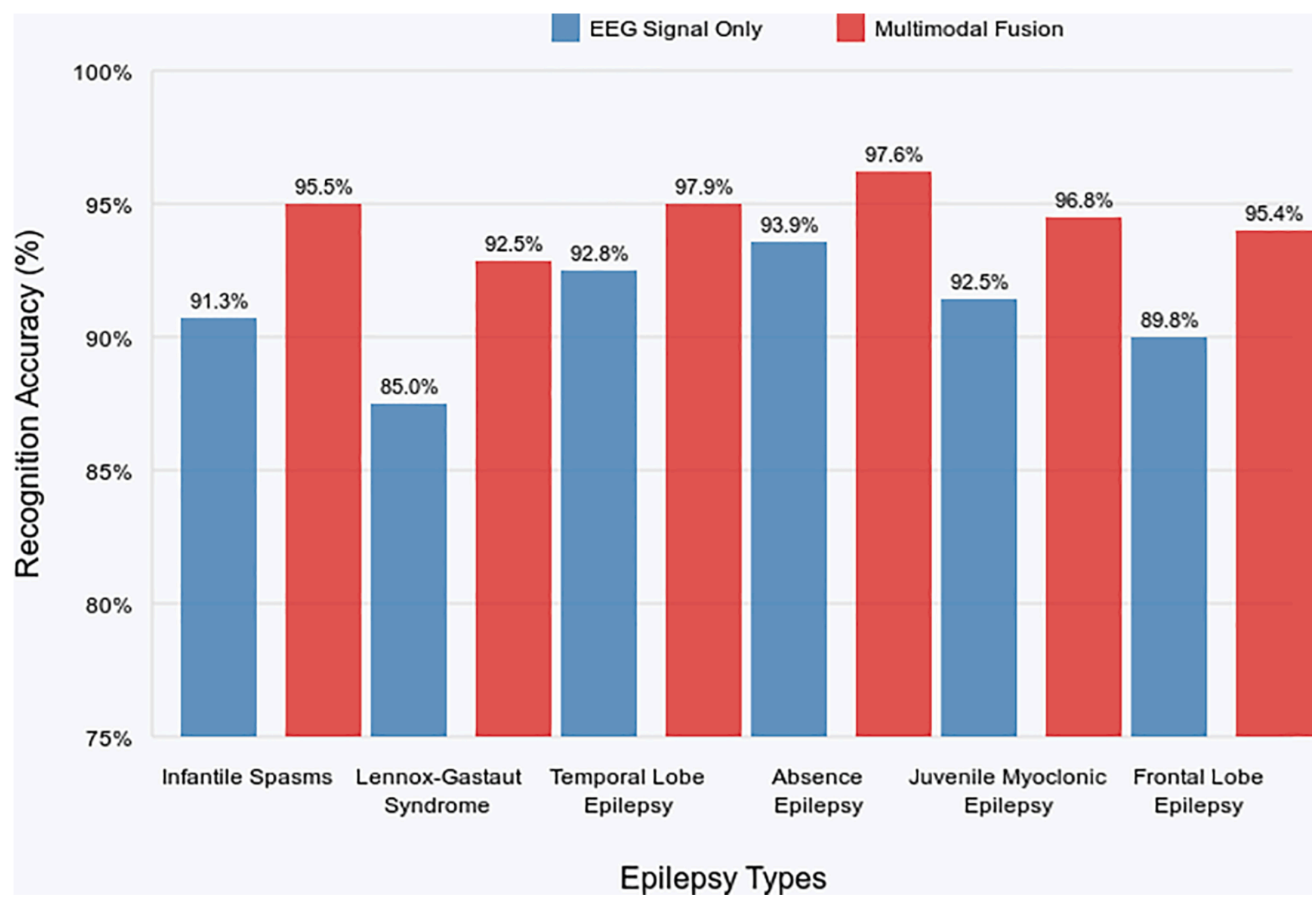

**Fig. 7.** Comparison of recognition accuracy of different epilepsy types using different input.

**Table 9**
Ablation experimental results of channel correlation discrimination switch mechanism.

| Configuration | Accuracy | Precision | Recall | F1 score | Calculation time |
|---|---|---|---|---|---|
| Complete System | 98.7% | 99.5% | 99.0% | 0.992 | 220ms |
| without the switch | 95.4% | 96.0% | 96.2% | 0.966 | 257ms |
| Performance changes | −3.3% | −3.5% | −2.8% | −0.026 | +37ms |

**Table 10**
The impact of the switch on epilepsy of different natures.

| Configuration | Accuracy of focal epilepsy identification | Accuracy of identifying generalized epilepsy |
|---|---|---|
| Complete System | 98.4% | 96.9% |
| without the switch | 95.5% | 95.9% |
| Performance changes | −2.9% | −1.0% |

Through these two sets of ablation experiments, this study verified the key contributions of multimodal data fusion and channel correlation discrimination switch mechanism to the performance of the main discrimination system, as depicted in Fig. 8. Multimodal fusion significantly improved the accuracy and early warning capability of the system, especially for the identification of complex epilepsy types; the channel correlation discrimination switch mechanism provided a more accurate processing strategy for the spatial characteristics of different types of epilepsy, while optimizing the computational efficiency.

### 4.2. Verification system performance

As a key component of the dual-system framework, the verification system not only provides independent epilepsy diagnosis results, but more importantly, it realizes the interpretable analysis of the diagnosis process through the SHAP method.

The verification system is based on feature engineering and random forest classifiers, and its overall performance on the test set is shown in Table 11.

The overall accuracy of the verification system is lower than that of the main discrimination system, which is expected because the verification system uses a simpler learning architecture. However, after automatic feedback learning, the accuracy of the verification system increased from 90.1% to 95.2%, and the F1 score increased from 0.898 to 0.916, which is a significant performance improvement, indicating that the feature weight adjustment mechanism effectively improves system performance.

The key advantage of the verification system is its interpretability. Through the feature importance score and SHAP analysis of the random forest classifier, this study can quantitatively evaluate the contribution of each feature to the diagnostic decision. Fig. 9 shows the change in feature importance ranking before and after automatic feedback learning.

As can be seen from Fig. 9 and Table 12, the automatic feedback learning mechanism significantly adjusted the feature importance ranking. In particular, the features related to delta (such as Ch1_DeltaWaveletEnergyRel and Ch1_DeltaRel) were significantly improved in ranking after feedback learning, indicating that the system correctly identified the key value of these features for epilepsy diagnosis, which is consistent with the phenomenon of abnormal enhancement of delta waves in EEG of epilepsy patients found in clinical studies.

At the same time, the ranking of the inter-channel phase lock value (PLV_ch1_ch2) has dropped, indicating that the system's reliance on spatial features has weakened, and it has focused more on time-frequency features. This dynamic adjustment of feature importance reflects that the automatic feedback learning mechanism can extract experience from high-quality predictions, continuously optimize feature weight allocation, and make the system closer to the diagnostic thinking of professional physicians.

The core advantage of the verification system lies in the diagnostic interpretability provided by the SHAP method. Fig. 10 shows the SHAP summary diagram, which intuitively presents the impact of each feature

**Table 11**
Performance indicators of the verification system on the test set.

| Indicators | Initial | After feedback | Improvements |
|---|---|---|---|
| Accuracy | 90.1% | 95.2% | +5.1% |
| Precision | 89.6% | 91.4% | +1.8% |
| Recall | 90.1% | 91.8% | +1.7% |
| F1 score | 0.898 | 0.916 | +0.018 |

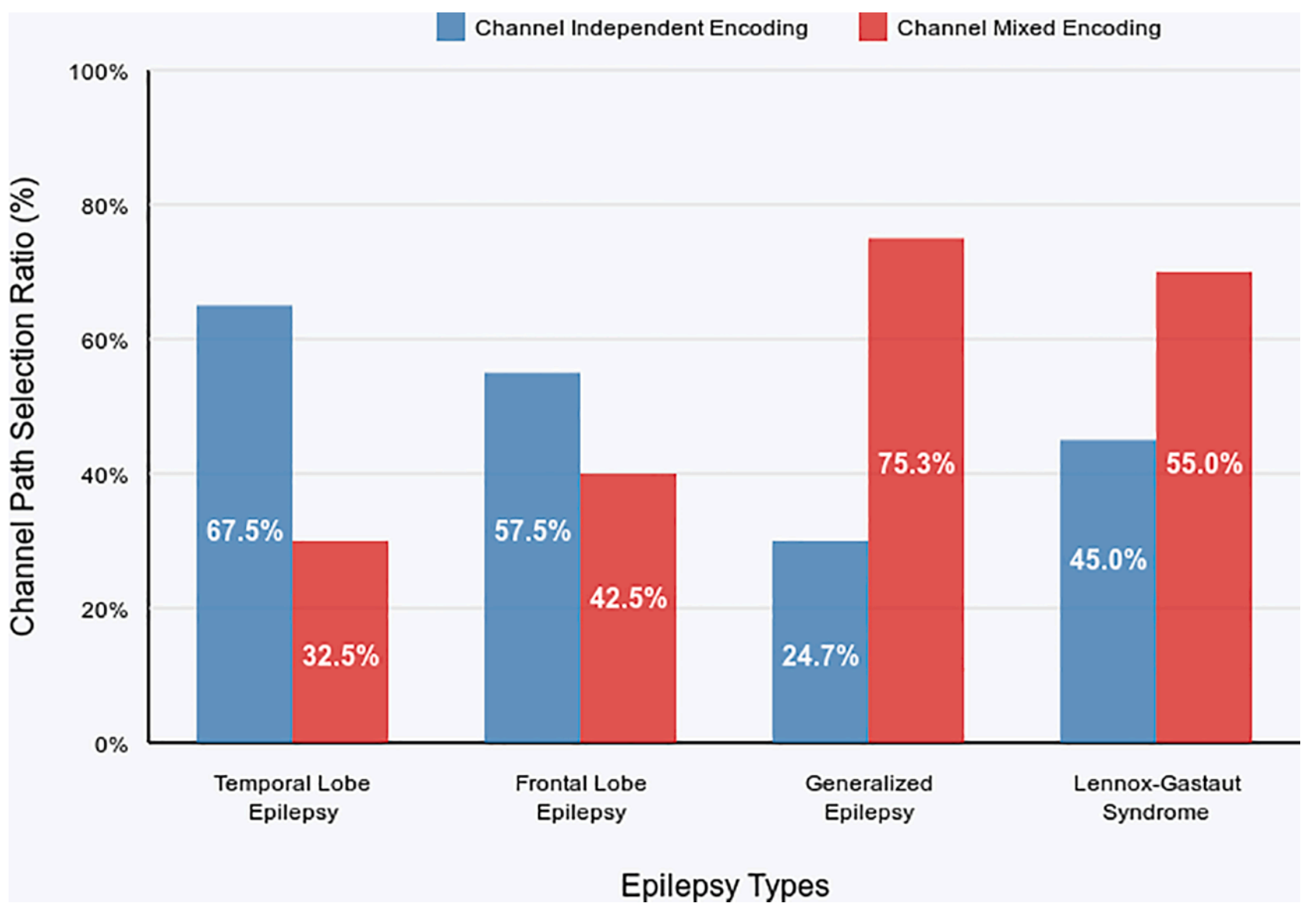

**Fig. 8.** The proportion of mixed channel selection in different epilepsy types.

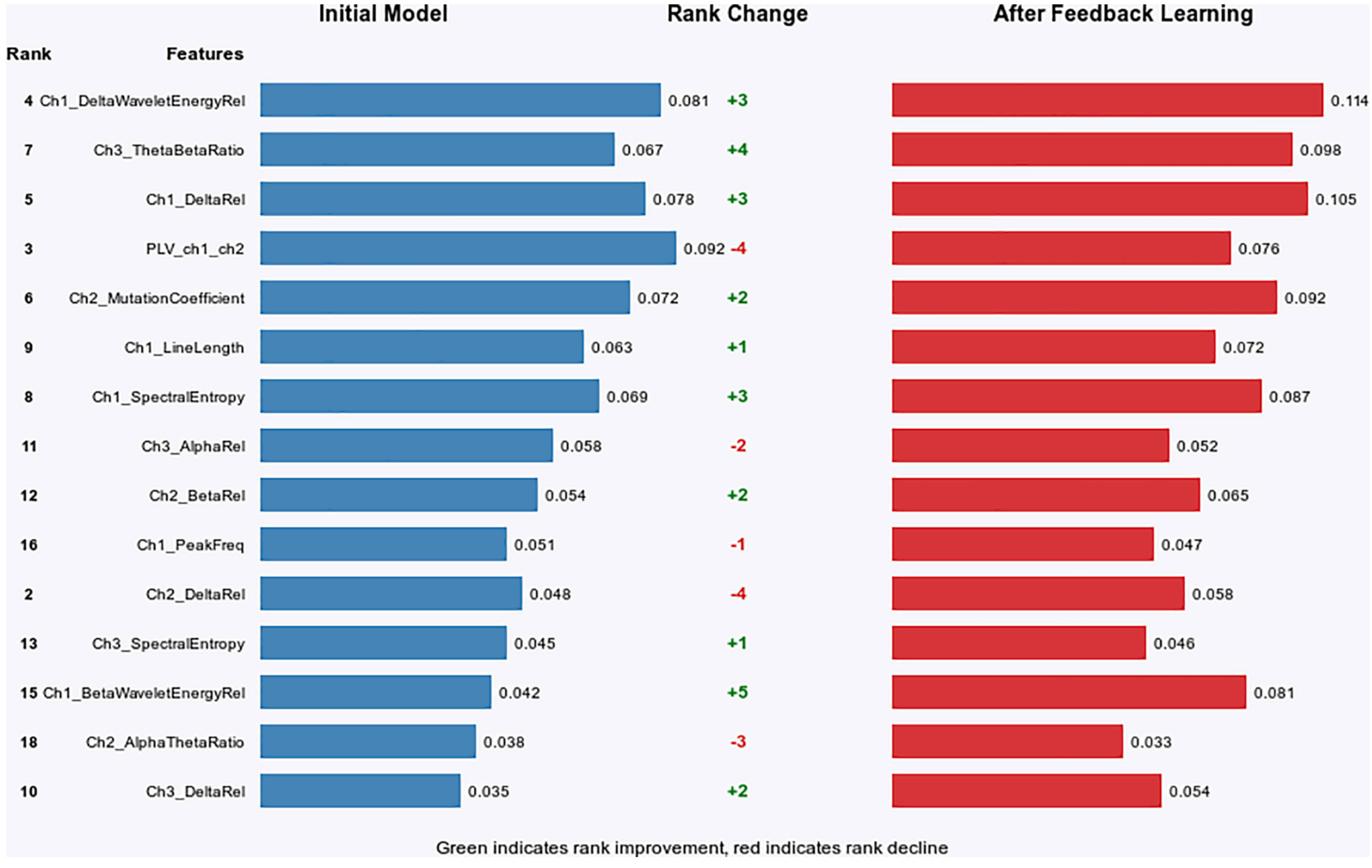

**Fig. 9.** Comparison of feature importance ranking before and after feedback.

**Table 12**
Top 10 features with the largest changes in feature importance ranking.

| Features | Initial ranking | Ranking after feedback | Ranking Changes |
|---|---|---|---|
| Ch1_DeltaWaveletEnergyRel | 4 | 1 | +3 |
| Ch3_ThetaBetaRatio | 7 | 3 | +4 |
| Ch1_DeltaRel | 5 | 2 | +3 |
| PLV_ch1_ch2 | 3 | 7 | −4 |
| Ch2_MutationCoefficient | 6 | 4 | +2 |
| Ch1_LineLength | 9 | 8 | +1 |
| Ch1_SpectralEntropy | 8 | 5 | +3 |
| Ch3_AlphaRel | 11 | 13 | −2 |
| Ch2_BetaRel | 12 | 10 | +2 |
| Ch1_PeakFreq | 16 | 17 | −1 |

on the prediction results.

It can be clearly observed from the SHAP summary diagram in Fig. 11 that Delta Wavelet Energy Rel contributes most to seizure prediction. High Delta Wavelet Energy (red dots) strongly pushes the prediction toward the seizure category, while low Delta Wavelet Energy (green dots) supports the normal state. This visual display enables doctors to understand the decision-making basis of the system and enhances trust in the AI system.

Particularly noteworthy is the importance of theta/beta ratio (ThetaBetaRatio). The figure shows that a high theta/beta ratio is highly correlated with epileptic seizures, which is consistent with the phenomenon of enhanced theta wave activity and beta wave suppression in epileptic patients found in neuroscience research. This consistency with clinical knowledge further verifies the reliability of the verification system.

Fig. 11 shows a SHAP waterfall chart that details the diagnostic path of a single sample. Starting from the baseline value (the average prediction of the model), the chart shows how each feature "pushes" the prediction result toward the final diagnosis. For example, in this epileptic seizure sample, the SHAP value of DeltaWaveletEnergyRel (+0.32) and the SHAP value of ThetaBetaRatio (+0.27) are key evidence supporting the diagnosis of seizures, while the SHAP value of AlphaRel (−0.14) suppresses this judgment. This detailed visual explanation provides doctors with a reference similar to a "second diagnostic opinion", making the AI diagnosis process transparent and understandable.

### 4.3. Dual system collaborative work effect

While dual-system agreement generally indicates higher reliability, we acknowledge that both systems could theoretically make the same error. Our analysis of failure cases provides important insights.

As shown in Table 13, when the two systems disagreed, manual review of the 415 disagreement cases revealed that the most common cause was complex partial seizures with subtle EEG changes (34%), followed by significant artefacts that affected one system disproportionately (28%), boundary cases between seizure types (22%), and rare seizure patterns not well-represented in training data (16%).

Analysis of false agreement cases revealed that among the 4967 agreement cases, 156 (3.1%) were incorrectly classified by both systems: 89 cases (57%) involved EEG artefacts mimicking seizure patterns, 42 cases (27%) were movement artefacts during behavioural seizures, and 25 cases (16%) involved rare seizure types with atypical presentations. This analysis demonstrates that while dual-system agreement generally improves reliability, systematic bias can still occur, particularly with unusual presentations or data quality issues.

The data shows that the dual systems reach high-confidence consistent diagnosis on 78.31% of the samples, and the accuracy of these samples is as high as 98.7%, close to the level of clinical experts. On 13.97% of the samples, the two systems reach low-confidence consistency with an accuracy of 98.1%. Only 6.72% show inconsistency between the systems, and the accuracy of these samples is significantly reduced to 92.0%. This result verifies the effectiveness of the dual-system framework - when the two systems diagnose consistently, the reliability of the results is significantly improved; when inconsistencies occur, the system can identify and mark these uncertain cases, reminding doctors to focus on them, and at the same time verifying the interpretability analysis of the system output as feedback to regulate the extraction of specific features by the main system and update the weights.

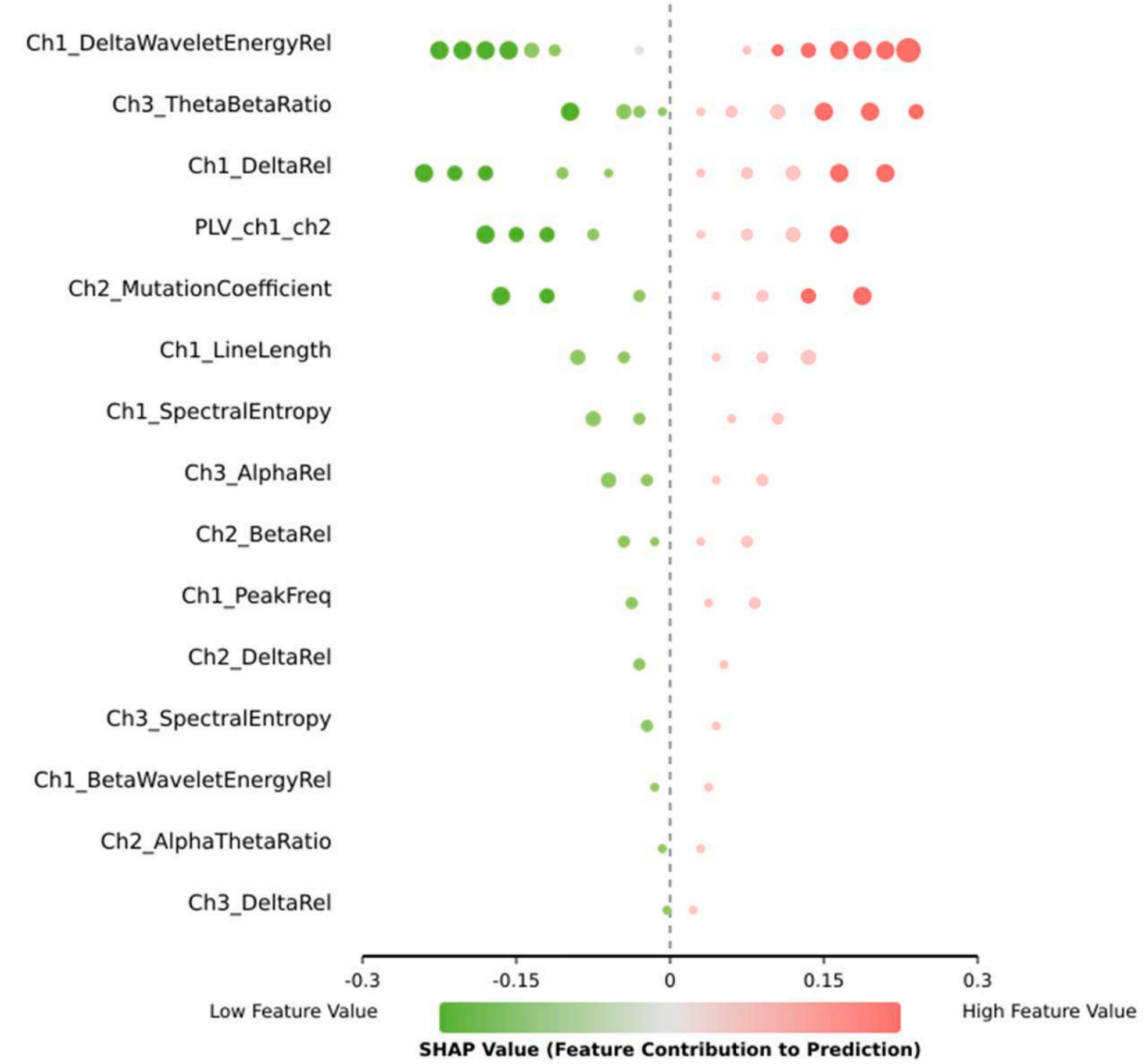

**Fig. 10.** Verification system SHAP summary diagram.

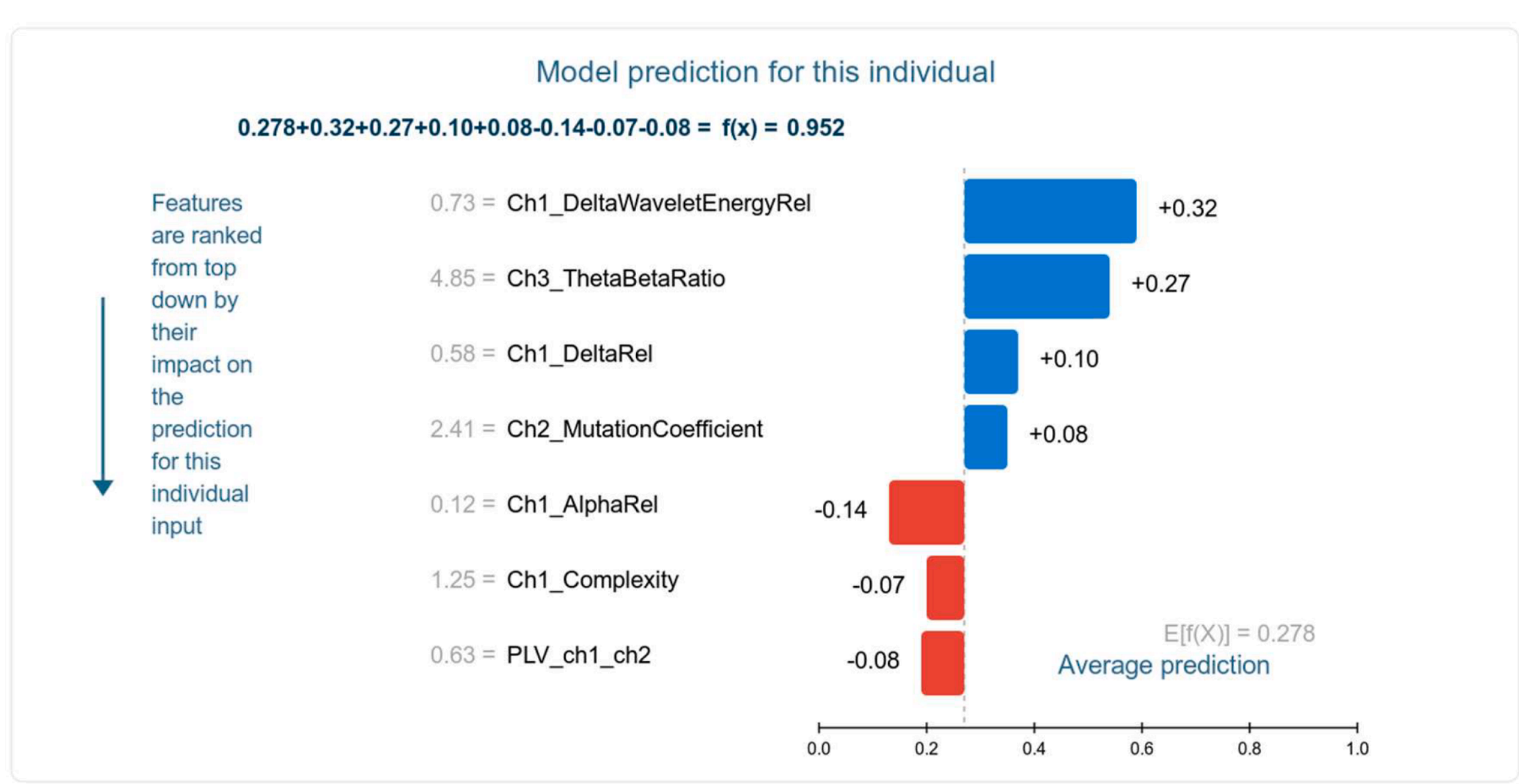

**Fig. 11.** SHAP waterfall plot of a single epileptic seizure sample.

The experimental results of inconsistent discrimination results output by the two systems also illustrate the important value of interpretability analysis of the verification system output in the task of intelligent diagnosis of epilepsy, proving that the single-mode end-to-end main system with a "black box" effect cannot output a diagnosis result that fully convinces doctors, so the collaborative work of the dual systems has its due meaning.

**Table 13**
Analysis of dual-system agreement patterns.

| Agreement Type | Sample Count | Percentage | Accuracy | Common Error Patterns |
|---|---|---|---|---|
| High-confidence agreement | 4215 | 78.31% | 98.7% | Rare false positives in noisy EEG segments |
| Low-confidence agreement | 752 | 13.97% | 98.1% | Boundary cases between interictal and ictal states |
| Disagreement cases | 415 | 7.72% | 92.0% | Complex seizure types, artifact contamination |

The core value of the dual-system collaborative work is to improve the diagnostic accuracy of the main discrimination system. Table 14 shows the performance of the main discrimination system after the collaborative mechanism.

Through the collaborative mechanism, the accuracy of the main discrimination system reached 98.7%, which is nearly 1 percentage point higher than the main discrimination system and about 5 percentage points higher than the verification system. The F1 score of the main system increased to 0.992, indicating that the system has achieved a better balance between accuracy and completeness.

### 4.4. Comparison with existing methods

In order to comprehensively evaluate the performance of the dual-system framework proposed in this study, this study compared it with existing epilepsy diagnosis methods. Table 15 shows the performance comparison of representative studies on the same or similar datasets.

As can be seen from Table 15, the dual-system framework of this study is superior to the existing best academic method in this field in terms of core indicators (accuracy and F1 score) of intelligent diagnostic systems - the SDCAE with Bi-LSTM method proposed by Mneimneh et al. in 2021. This also reflects to a certain extent that the multi-source and multi-modal joint data set contributes to the accuracy of the diagnostic results output by the main system. In addition, the method of Mneimneh et al. lacks in-depth interpretability analysis, while the method of this study uses the SHAP method for feature contribution analysis, which greatly enhances the transparency of the model decision-making process and provides higher credibility for clinical applications.

We acknowledge a significant limitation of our study: all experiments were conducted on our integrated dataset without external validation on completely independent datasets from different institutions.

This limitation affects the generalizability claims of our approach in several ways. Our model may have learned patterns specific to the recording equipment, protocols, or patient populations represented in our training data, creating institution-specific biases that could limit performance when applied to other clinical settings.

Additionally, the extensive preprocessing pipeline we developed may not generalize to data collected under different conditions or with different equipment specifications, as variations in recording parameters, electrode placement precision, or signal processing standards could impact model performance.

Furthermore, the demographic and clinical characteristics of patients in our dataset may not represent the broader epilepsy population globally, potentially limiting the applicability of our findings across diverse patient groups with different genetic backgrounds, seizure etiologies, or clinical presentations.

**Table 14**
Dual-system synergy improves dual-system performance.

| Indicators | The main system | The verification system |
|---|---|---|
| Accuracy | +0.9% | +5.1% |
| Precision | +0.8% | +1.8% |
| Recall | +0.9% | +1.7% |
| F1 score | +0.008 | +0.018 |

To address these limitations, we propose a comprehensive validation strategy that involves collaboration with additional medical centers to obtain independent test datasets from institutions with different recording protocols and patient populations. This external validation would include systematic evaluation of model performance across different EEG recording systems and protocols to assess robustness to technical variations, as well as assessment of performance across diverse patient populations and epilepsy types to ensure broad clinical applicability.

We also plan to develop domain adaptation techniques that can improve cross-institutional generalizability by automatically adjusting model parameters to account for systematic differences between recording environments and patient populations, thereby enhancing the real-world deployment potential of our dual-system diagnostic framework.

Based on the above analysis, the dual-system framework of this study achieves an effective balance between high accuracy and high interpretability through the collaborative work of the main discrimination system and the verification system.

In terms of computational efficiency, as shown in Table 16, when processing EEG sequences of the same length and complexity (including fractal dimension), the computational time of this research method is significantly lower than that of other methods based on academia, and it also consumes less memory and has fewer floating-point operations per second.

The computational efficiency advantage of the Mamba-Bi-LSTM architecture mainly stems from the linear computational complexity of the Mamba model. When processing a 60-second EEG sequence, the computation time of the New Mamba with Bi-LSTM method is about 68% faster than that of the SDCAE with Bi-LSTM method, and the memory consumption is reduced by about 61%. This high efficiency makes this research method more suitable for resource-constrained clinical deployment environments, such as portable EEG monitoring devices or mobile medical platforms.

In summary, the dual-system framework proposed in this study is superior to existing methods in multiple dimensions such as accuracy, interpretability, and computational efficiency, providing a more comprehensive solution for intelligent auxiliary diagnosis of epilepsy.

## 5. Discussion

Our dual-system framework for epilepsy diagnosis addresses three requirements that have rarely been achieved simultaneously in prior work: high diagnostic accuracy, computational efficiency sufficient for real-time clinical deployment, and decision-level transparency that supports rather than replaces physician judgment. The experimental results demonstrate that these goals are not in tension. Each architectural choice reinforces the others and the framework achieves them consistently across five heterogeneous datasets spanning different recording modalities, classification tasks, and clinical contexts.

The performance of the main discrimination system warrants careful interpretation beyond the headline accuracy figure. Our overall accuracy of 98.7% with an F1 score of 0.992 represents a 25% reduction in diagnostic errors compared to the previous best method — from 94 misclassifications to 70 across our 5382-sample test set. This improvement stems from three complementary architectural innovations. The Mamba module's selective state-space design achieves linear computational complexity with respect to sequence length, enabling effective modelling of the long EEG recordings characteristic of clinical monitoring without the quadratic cost of self-attention. The multimodal fusion strategy captures complementary information from EEG signals and clinical text records, with particularly pronounced benefits for complex syndromes such as Lennox-Gastaut syndrome, where EEG alone is insufficient to capture the full diagnostic picture. The channel correlation discrimination switch mechanism adaptively routes processing through channel-independent or channel-mixed encoding paths depending on the inter-channel correlation structure, which aligns directly with the pathophysiological distinction between focal seizures — originating from spatially restricted brain regions — and generalised seizures involving synchronous widespread discharge.

**Table 15**
Performance comparison between this study and existing epilepsy diagnosis methods.

| Methods | Author and year | Dataset | Accuracy | F1 score | Interpretability |
|---|---|---|---|---|---|
| New Mamba with Bi-LSTM | Mufeng Chen, 2025 | University of Bonn Dataset, CHB-MIT, EPILEPSIAE TUH Corpus | 98.7% | 0.992 | Feature contribution analysis based on SHAP method |
| SDCAE with Bi-LSTM[14] | Agwad, Radu, et al., 2021 | CHB-MIT | 98.26% | 0.9826 | None |
| LSTM[1] | Lipton, Zachary C., et al.,2015 | UCI-Epileptic Seizure Recognition Dataset | 97% | 0.93 | None |
| XGBoost[3] | Vanabelle, Pierre, et al.,2020 | TUH EEG Seizure Corpus | 77% | 0.51 | GBoost's decision tree feature, estimating feature importance |
| Two-block algorithm with channel annotation[4] | Wong, Sheng, et al., 2025 | TUH EEG Seizure Corpus | 88% | — | Channel attention mechanism , transparent view of the decision process |
| Machine-learning-based diagnostics of EEG pathology[10] | Gemein, Lukas A.W., et al.,2020 | TUH Abnormal EEG Corpus | 81%–86% | — | Feature Importance Analysis |
| ResBiLSTM[21] | Zhao et al., 2024 | Bonn, TUSZ | 95.0% | 0.950 | Not mentioned |
| Multidimensional Transformer + LSTM-GRU[22] | Li et al., 2024 | Bonn, CHB-MIT | ~98% | 0.987 | Not mentioned |
| 1D CNN-LSTM + DWT[23] | Sun et al., 2025 | Bonn, CHB-MIT, TUSZ | 97.24% | — | Not mentioned |
| Modified Transformer (Inresformer) + DWT[24] | Hu et al., 2025 | Bonn, CHB-MIT | 98.03% | — | Not mentioned |

**Table 16**
Comparison of computational efficiency of processing the same 60-second EEG sequences with typical methods.

| Methods | Author and year | Calculation time | Memory consumption |
|---|---|---|---|
| New Mamba with Bi-LSTM | Mufeng Chen, 2025 | 220ms | 485MB |
| SDCAE with Bi-LSTM | Agwad, Radu, et al., 2021 | 687ms | 1250MB |
| LSTM | Lipton, Zachary C., et al.,2015 | 512ms | 860MB |
| XGBoost | Vanabelle, Pierre, et al.,2020 | 342ms | 620MB |
| Two-block algorithm with channel annotation | Wong, Sheng, et al., 2025 | 331ms | 578MB |
| Machine-learning-based diagnostics of EEG pathology | Gemein, LukasA.W., et al.,2020 | 569ms | 896MB |

What distinguishes our work most sharply from the existing literature is not accuracy alone, but the combination of capabilities that no single prior method provides. Most published methods are evaluated on a single benchmark dataset and do not report early warning time as a performance metric. Our framework achieves a 15-second advance warning window, extended by 8.8 s over the EEG-only configuration, which has direct clinical significance: 15 s is sufficient to activate neurostimulation devices, guide patients to safe positions, and alert caregivers, whereas the 2–6 s warning typical of single-modality approaches is often too brief for meaningful intervention. Furthermore, virtually all high-performing methods in the literature operate as end-to-end black-box systems. Recent work by Hu et al. [24] achieved 100% accuracy on the Bonn dataset and 98.03% on CHB-MIT using a modified Transformer with discrete wavelet transform preprocessing, and Sun et al. [23] reported strong and consistent performance across Bonn, CHB-MIT, and TUH simultaneously with a 1D CNN-LSTM architecture, yet neither provides a mechanism for clinicians to understand or validate individual predictions. In contrast, our SHAP-based verification system is not a post-hoc explanation appended to an existing model; it is an architecturally independent diagnostic component that participates directly in the final decision, cross-validates the main system's outputs, and flags disagreements for human review. This functional role — active verification rather than passive explanation — represents a structural departure from how interpretability has been treated in the epilepsy AI literature.

The per-dataset results reveal further nuance about the framework's robustness. Performance ranges from 96.4% on the heterogeneous TUH Corpus to 99.2% on the high-quality EPILEPSIAE intracranial recordings, with surface EEG datasets (Bonn, CHB-MIT, Karunya) clustering in the 97.8–98.9% range. This 2.8 percentage point spread across datasets with fundamentally different signal characteristics — invasive versus non-invasive recording, binary versus multi-class tasks, controlled laboratory conditions versus real-world clinical noise — is considerably narrower than the variation typically observed when applying a single-dataset-optimised method to out-of-distribution data. Intracranial EEG naturally provides superior signal quality due to electrode proximity to neural sources, which explains EPILEPSIAE's peak performance, while the TUH Corpus's real-world clinical complexity — encompassing diverse patient populations, variable recording conditions, and naturally occurring artefacts — presents a more challenging classification problem. Critically, these results enable direct comparison with published benchmarks (Table 15): on the Bonn three-class task, we achieve 98.9%, surpassing recent benchmarks including the ResBiLSTM method of Zhao et al. [21] (95.0%) and the 1D CNN-LSTM approach of Sun et al. [23] (97.24%). On the challenging TUH Corpus, our 96.4% substantially exceeds the benchmarks of Wong et al. [4] (88%) and Gemein et al. [10] (81–86%), an improvement of 8–15 percentage points on the most clinically realistic dataset in our evaluation. For CHB-MIT, we achieve 97.8% compared to 98.26% for the dataset-specific optimised method of Agwad et al. [14] — a gap of only 0.5 percentage points despite our model being trained on five datasets simultaneously, demonstrating that generalisability does not require sacrificing specialised performance. This pattern of consistent high performance across heterogeneous datasets provides strong evidence that the Mamba-Bi-LSTM architecture and multimodal fusion strategy capture fundamental epileptic patterns that generalise across recording modalities and clinical contexts, rather than overfitting to the characteristics of any single data source.

The computational profile of the system is equally important for clinical translation. At 220 ms per 60-second EEG segment with 485 MB memory consumption, our system operates 68% faster and requires 61% less memory than comparable deep learning approaches, as shown in Table 16. This efficiency is a direct consequence of the Mamba architecture's linear scaling, and it enables deployment on standard clinical workstations and resource-constrained portable monitoring devices without specialised hardware acceleration. The SHAP-based verification adds negligible latency to this pipeline given its feature-engineering foundation, maintaining overall system responsiveness well within the 500 ms threshold required for real-time clinical use.

The automatic feedback mechanism deserves particular attention as a source of continuing performance improvement. High-confidence agreements between the main and verification systems are recycled as training signal, refining both the neural architecture's weights and the verification system's feature importance rankings. This creates a virtuous cycle: the main system's confident predictions guide the verification system's feature prioritisation, which in turn highlights diagnostically relevant signal characteristics for the main system's ongoing learning. The result is a 5.1% accuracy improvement in the verification system and a 0.9% improvement in the main system following feedback integration, with diagnostic consistency achieved on 92% of test cases. The 8% of cases flagged as disagreements are precisely the ambiguous or artefact-affected samples where independent human review is most warranted — the system's uncertainty is, in this sense, calibrated and clinically informative rather than arbitrary.

It is appropriate to acknowledge the field's ongoing rapid development. A very recent study by Jeong et al. (2026) demonstrated real-time seizure detection and multi-seizure classification on paediatric EEG using a ResNet-LSTM hybrid architecture, achieving an AUROC of 0.98 on an independent clinical dataset from Severance Children's Hospital [25]. While their dataset differs from the standard benchmarks used in our evaluation — and no peer-reviewed primary research articles published in 2026 to our knowledge yet report quantitative results on CHB-MIT, Bonn, or TUH in a manner directly comparable to our work — this study reinforces the growing clinical demand for real-time, multi-class capable systems. It also highlights interpretability as the next critical frontier: like most high-performing architectures in the current literature, the Jeong et al. system does not provide prediction-level explanations accessible to clinicians, a gap that our dual-system design was specifically built to address.

Several limitations of the present work must be acknowledged. All experiments were conducted on our integrated multi-institutional dataset; external validation on data from institutions not represented in training remains the most important unresolved question for clinical translation. The model may have internalised institution-specific patterns — related to recording equipment, electrode placement protocols, or patient population characteristics — that do not generalise to settings with different technical standards. The preprocessing pipeline, while systematic, introduces its own source of variation, and slight differences in artefact removal or signal normalisation could affect downstream performance in ways that are difficult to anticipate without prospective testing in new environments. The verification system, despite its interpretability advantage, remains data-driven and cannot enforce compliance with established clinical diagnostic criteria or ILAE classification guidelines as hard constraints. Finally, both systems share susceptibility to systematic failure when encountering artefact patterns that consistently mimic seizure morphology across modalities, a vulnerability that the dual-system architecture mitigates but does not eliminate.

These limitations define the agenda for future work. Multi-institutional prospective validation is the most urgent priority, with federated learning offering a path to training on diverse institutional data while preserving patient privacy. The performance differential between intracranial and surface EEG recordings (99.2% versus 96.4–98.9%) warrants targeted domain adaptation work, potentially through transfer learning strategies that leverage the higher signal quality of intracranial recordings to improve surface EEG models. Integrating explicit medical knowledge, through hybrid architectures that combine learned representations with rule-based reasoning grounded in clinical ontologies, would address the gap between statistical performance and clinical guideline compliance. Extensions to finer-grained seizure subtype classification, longitudinal personalisation of model parameters to individual patient trajectories, and integration with other neuroimaging modalities represent natural directions for expanding clinical utility.

The broader significance of this work lies in its demonstration that accuracy and interpretability are complementary rather than competing objectives in medical AI. The dual-system framework shows that embedding an independent verification component, one that both explains and cross-validates, can improve overall system reliability while simultaneously addressing the transparency requirements that are increasingly central to regulatory frameworks and clinical adoption standards. As AI systems take on more prominent roles in high-stakes diagnostic workflows, this architectural principle — designing for explainability as a functional requirement rather than an afterthought — may prove as important as the specific technical advances reported here.6. Conclusion

In this study, we successfully developed a medical AI-assisted diagnosis system based on deep learning for epilepsy, a disease that seriously threatens the life and health of patients.In order to address the difficulty of balancing accuracy and interpretability in clinical diagnosis, we proposed a dual-system intelligent diagnosis framework with a main discrimination system and a verification system in parallel, which increased the balance between diagnostic accuracy and explanatory transparency.

The experimental results show that the epilepsy intelligent diagnosis system achieved an accuracy of 98.7% and an F1 score of 0.992, and the core indicators were better than the best existing methods in academia. The effective integration of key technologies such as multimodal data fusion, channel correlation processing, and an efficient hybrid network built by combining the Mamba architecture with Bi-LSTM enables the system to demonstrate good adaptability and stability in a variety of medical scenarios. Our key contributions include the novel Mamba-Bi-LSTM hybrid architecture achieving linear computational complexity, comprehensive multimodal data fusion demonstrating superior performance across epilepsy types, and rigorous evaluation methodology with detailed ablation studies and failure case analysis. The system's 15-second early warning capability and SHAP-based interpretability enhance both clinical value and physician trust.

However, we acknowledge important limitations: external validation on independent datasets is required to confirm generalizability, the verification system still relies on data-driven methods rather than explicit domain knowledge, and performance may vary across different recording environments and patient populations.

In summary, the dual-system intelligent diagnosis framework developed in this study provides a new technical route and methodological reference for the field of medical AI-assisted diagnosis, while achieving high-precision diagnosis and maintaining the transparency of the decision-making process. Future research will focus on addressing these limitations through multi-institutional validation studies and integration of explicit medical knowledge into the verification framework, improving the system's generalization ability, lightweight deployment and clinical integration, thereby promoting medical AI technology to develop in a more accurate, reliable and inclusive direction.

All dataset annotations were created and validated by the original institutions; our study did not involve any human annotation activity.

## CRediT authorship contribution statement

**Mufeng Chen:** Writing – review & editing, Writing – original draft, Visualization, Validation, Supervision, Software, Resources, Project

administration, Methodology, Investigation, Formal analysis, Data curation, Conceptualization. **Jia Xie:** Visualization, Software, Resources, Project administration, Formal analysis, Conceptualization. **Fuchang Luo:** Writing – original draft, Visualization, Supervision, Software, Resources, Methodology, Investigation, Formal analysis, Conceptualization. **Quansheng Ren:** Writing – review & editing, Validation, Resources, Project administration, Funding acquisition, Formal analysis, Data curation, Conceptualization.

## Declaration of competing interest

The authors declare that they have no known competing financial interests or personal relationships that could have appeared to influence the work reported in this paper.

## Acknowledgments

The work was supported by Beijing Natural Science Foundation (L248094), and supported in part by the High Performance Computing Platform of Peking University.

## Data availability

Data will be made available on request.